\documentclass[12pt]{JHEP3}

\pdfoutput=1

\usepackage{amssymb,amsmath,amsfonts,epsfig, subfigure}
\usepackage{cite}
\usepackage{slashed}
\usepackage{picture,epic}

\newcommand{\beqs}{\begin{equation*}}
\newcommand{\beq}{\begin{equation}}

\newcommand{\eeqs}{\end{equation*}}
\newcommand{\eeq}{\end{equation}}

\newcommand{\beqas}{\begin{eqnarray*}}
\newcommand{\beqa}{\begin{eqnarray}}

\newcommand{\eeqas}{\end{eqnarray*}}
\newcommand{\eeqa}{\end{eqnarray}}

\newcommand{\eq}[2]{\begin{equation} #1 \label{#2} \end{equation}}

\newcommand{\eps}{\varepsilon}
\newcommand{\al}{\alpha}
\newcommand{\be}{\beta}
\newcommand{\ga}{\gamma}
\newcommand{\de}{\delta}

\newcommand{\ka}{\kappa}
\newcommand{\la}{\lambda}
\newcommand{\si}{\sigma}

\newcommand{\Ga}{\Gamma}
\newcommand{\De}{\Delta}

\newcommand{\La}{\Lambda}

\newcommand{\blist}{\begin{itemize}}

\newcommand{\elist}{\end{itemize}}

\providecommand{\href}[2]{#2}

\DeclareFontFamily{OT1}{rsfs}{}
\DeclareFontShape{OT1}{rsfs}{m}{n}{ <-7> rsfs5 <7-10> rsfs7 <10->rsfs10}{} 
\DeclareMathAlphabet{\mycal}{OT1}{rsfs}{m}{n}

\DeclareMathOperator{\extdm}{d}
\newcommand{\extd}{\extdm \!}

\newcommand{\cO}{\mathcal{O}}
\newcommand{\cD}{\mathcal{D}}
\newcommand{\cH}{\mathcal{H}}
\newcommand{\vecm}{{\bf m}}
\newcommand{\Osq}{{\cal O}^{\rm log^2}}
\newcommand{\Lie}{{\cal L}}

\title{Short-cut to new anomalies in gravity duals to logarithmic conformal field theories}

\author{Daniel Grumiller, Niklas Johansson and Thomas Zojer\\
           Institute for Theoretical Physics, 
           Vienna University of Technology,\\
           Wiedner Hauptstr. 8--10/136,
           A-1040 Vienna, Austria\\
           E-mails: \email{grumil@hep.itp.tuwien.ac.at, niklasj@hep.itp.tuwien.ac.at, e0526703@student.tuwien.ac.at}}

\abstract{
Various massive gravity theories in three dimensions are conjecturally
dual to logarithmic conformal field theories (LCFTs).
We summarise the status of these conjectures. LCFTs are characterised
by the values of the central charges and the so-called
``new anomalies''. We employ a short-cut
to calculate these new anomalies in generalised massive gravity and in the recently proposed
higher-derivative gravity theories with holographic $c$-theorem. Both cases permit LCFTs exhibiting intriguing features, like rank three Jordan cells or non-zero central charges. 
Finally, as an example we discuss in some
detail the partially massless version of new massive
gravity, a theory with several special properties that we call ``partially massless gravity''.
}


\preprint{TUW--10--13}

\begin{document}

\section{Introduction}

Logarithmic conformal field theories (LCFTs) are conformal field theories (CFTs) where correlation functions and operator product expansions may contain logarithms \cite{Gurarie:1993xq}. LCFTs arise in various contexts in condensed matter physics, see the introductions of \cite{Flohr:2001zs,Gaberdiel:2001tr}. A defining feature of LCFTs is that the Hamiltonian does not diagonalise, but rather contains Jordan cells of rank two or higher. Thus, there are (at least) two operators with degenerate conformal dimensions, which can have non-trivial correlators with each other. Another relevant difference to ordinary CFTs is that LCFTs are not unitary.

Essential features of CFTs are characterised by the values of the central charges $c_{L/R}$: the anomalous term in the conformal Ward identities, the anomalous term in the transformation of the energy-momentum tensor, the number of microstates counted via the Cardy formula etc. Consequently, there are many independent ways to calculate the central charges. Depending on the context, some of them may be considerably simpler than others. 

Essential features of LCFTs are characterised by the values of the central charges {\em and} the values of so-called ``new anomalies''. The main purpose of our work is to explain and exemplify 
an efficient way to calculate on the gravity side the values of new anomalies for LCFTs that have gravity duals.

For sake of concreteness we focus first on a very specific class of LCFTs, namely one where the energy-momentum tensor degenerates with another operator, its logarithmic partner, and where the corresponding central charge vanishes. Such LCFTs arise for instance in condensed matter physics systems at (or near) a critical point with quenched disorder, like spin glasses \cite{Binder:1986zz}/quenched random magnets \cite{Cardy:1999zp,RezaRahimiTabar:2000qr}, dilute self-avoiding polymers or percolation \cite{Gurarie:1999yx}. We denote the (anti-)holomorphic flux components of the energy-momentum tensor by ${\cal O}^L(z)$ (${\cal O}^R(\bar z)$). Let us suppose that the holomorphic flux component ${\cal O}^L(z)$ acquires a logarithmic partner ${\cal O}^{\rm log}(z,\bar z)$. Then the non-vanishing 2-point correlators are given by\footnote{We omit terms that are less divergent than the anomalous term in the near coincidence limit $z,\bar z\to 0$ as well as contact terms.}
\begin{subequations}
\label{eq:short1}
\begin{align}
& \langle{\cal O}^R(\bar z)\,{\cal O}^R(0)\rangle = \frac{c_R}{2\bar z^4} \\
& \langle{\cal O}^L(z){\cal O}^{\rm log}(0,0)\rangle = \frac{b_L}{2z^4} \label{eq:Llog} \\
& \langle{\cal O}^{\rm log}(z,\bar z){\cal O}^{\rm log}(0,0)\rangle = -\frac{b_L \ln{(m^2_L|z|^2)}}{z^4} \label{eq:loglog}
\end{align}
\end{subequations}
The left central charge vanishes, $c_L=0$. 
The quantity $b_L$ is the new anomaly. 
The mass scale $m_L$ is spurious and can be changed to any finite value by the redefinition ${\cal O}^{\rm log}\to{\cal O}^{\rm log}+\ga{\cal O}^L$, with some finite $\ga$. Since the ${\cal O}^L$ 2-point correlator vanishes, the new anomaly is well-defined only after an over-all normalisation for the pair ${\cal O}^L$ and ${\cal O}^{\rm log}$ has been fixed.

For LCFTs with the correlators \eqref{eq:short1} above the Hamiltonian $H$ is not diagonalisable.
Rather, it acquires a rank 2 Jordan cell.
\eq{ 
H \left(\begin{array}{c} {\cal O}^{\rm log} \\ {\cal O}^L
\end{array}\right) = \left(\begin{array}{c@{\quad}c}
2 & 1 \\
0 & 2
\end{array}\right) \left(\begin{array}{c} {\cal O}^{\rm log} \\ {\cal O}^L \end{array}\right)
}{eq:cg79} 
Consistency of the LCFT, in particular locality, requires that the angular momentum $J$ is diagonalisable.
\eq{ 
J \left(\begin{array}{c} {\cal O}^{\rm log} \\ {\cal O}^L
\end{array}\right) = \left(\begin{array}{c@{\quad}c}
2 & 0 \\
0 & 2
\end{array}\right) \left(\begin{array}{c} {\cal O}^{\rm log} \\ {\cal O}^L \end{array}\right)
}{eq:cg1} 
The eigenvalues $2$ arise because the energy-momentum tensor and its logarithmic partner both correspond holographically to spin-2 excitations.

We turn now to the gravity side of LCFTs.
It is of course not at all clear that there is a gravity side in the first place: usual applications of the AdS/CFT correspondence \cite{Maldacena:1997re} 
involve unitary theories on the gravity side (string theory) and on the field theory side (unitary CFTs).
Nevertheless, it is a logical possibility that there exist non-unitary gravitational theories that can serve as duals to LCFTs.
It was conjectured in \cite{Grumiller:2008qz} that topologically massive gravity (TMG) \cite{Deser:1982vy,
Deser:1982sv} for a certain tuning of its parameters is dual to an LCFT with the properties above, \eqref{eq:short1}-\eqref{eq:cg1}.
We review the status of this conjecture and summarise the evidence in the next section. 
We just mention here that the conjecture originally was based upon the observation that the defining properties \eqref{eq:cg79}-\eqref{eq:cg1} hold.
Let us therefore suppose as a reasonable working hypothesis that the LCFT conjecture is true for TMG.
Then an obvious question to ask is: what is the value of the new anomaly $b_L$?
This question was answered by brute-force calculations of 2-point correlators \cite{Skenderis:2009nt} and 2-, 3-point correlators \cite{Grumiller:2009mw} on the gravity side.
Skenderis, Taylor and van Rees found additionally a short-cut to cross-check their result for the new anomaly \cite{Skenderis:2009nt} that exploits well-known LCFT limiting constructions, see e.g.~\cite{Rasmussen:2004gx, Rasmussen:2004na}.
While it is always nice to have shorter calculations, there would be no urgent need to elaborate on this short-cut 
to new anomalies if TMG was the only example of an LCFT dual.

Interestingly, it turned out that TMG is not the only example that could serve as a gravity dual to an LCFT. 
New massive gravity (NMG) \cite{Bergshoeff:2009hq,Bergshoeff:2009aq} also allows for a tuning of the coupling constants such that an LCFT emerges, albeit with properties that differ slightly from the ones above \eqref{eq:short1}-\eqref{eq:cg1}.
Again, the new anomalies were determined by brute-force calculations of 2-point correlators \cite{Grumiller:2009sn,Alishahiha:2010bw}. 
In complete analogy to \cite{Skenderis:2009nt}, it was noted in \cite{Grumiller:2009sn} that there is a short-cut to determine the new anomalies that avoids the calculation of 2-point correlators. 
This short-cut is what we flesh out in the present work, with the aim to apply it to more general massive 
gravity theories in three dimensions.


There are two ways an LCFT can occur in NMG. As opposed to TMG, NMG is parity invariant, so when a type of LCFT described by \eqref{eq:short1}-\eqref{eq:cg1} emerges actually both $\cO^L$ and $\cO^R$ degenerate with a separate massive mode, each resulting in a logarithmic pair in a theory with $c_L = c_R = 0$. The other possibility is that the two massive modes degenerate with each other, leading to a theory with $c_L = c_R \neq 0$ and exhibiting many, not unrelated, special features: the Breitenlohner--Freedman bound is saturated, normalisable and non-normalisable modes mix, the modes exhibit a new kind of asymptotic behaviour, and display partial masslessness in the sense of Deser and Waldron \cite{Deser:1983mm,Deser:2001pe} 
resulting in gauge enhancement. We shall explore this theory --- an example for ``partially massless gravity'' --- in some detail in the present work. 

Combining TMG and NMG yields yet-another 3-dimensional massive gravity theory, generalised massive gravity (GMG).
Given that both TMG and NMG can serve as gravity duals for specific LCFTs it is natural to conjecture that the same applies to GMG.
The new anomalies for GMG have not been calculated yet.
Since GMG has an additional coupling constant as compared to TMG or NMG there is a richer spectrum of dual LCFTs.
In particular, for the first time Jordan cells of rank 3 can emerge.
Moreover, when two massive modes degenerate it is possible to obtain LCFTs where both central charges are non-vanishing and unequal. Thus, it is of interest to study GMG and its LCFT duals in detail.

A first step in this direction is to answer the obvious question about the values of new anomalies.
In this work we use the above mentioned short-cut to calculate the new anomalies for GMG.
This short-cut takes advantage of the fact that the LCFTs in question arise as limits of (non-unitary) CFTs.
Therefore it is possible to infer the new anomaly from the behaviour of the correlators and weights of the 
operators in question as this limit is taken. For the case \eqref{eq:short1}-\eqref{eq:cg1} where the correlators are
determined by the central charges, this information is readily available on the gravity side. For the case when two massive modes degenerate the 2-point correlator of these modes is needed to use the short-cut.
This method also works for the infinite set of higher derivative extensions of NMG consistent with a holographic $c$-theorem \cite{Sinha:2010ai,Paulos:2010ke,Sinha:2010pm}. 

This paper is organised as follows: in Section \ref{sec:2} we review the status of the LCFT conjecture for TMG. 
In Section \ref{sec:3} we formulate the LCFT conjecture for NMG and GMG and collect evidence in its favour.
In Section \ref{sec:4} we explain the short-cut to determine the new anomalies and apply it to TMG and NMG.
In Section \ref{sec:5} we apply the short-cut to determine the new anomalies for GMG.
In Section \ref{sec:6} we study an example for partially massless gravity.
In Section \ref{sec:7} we address generalisations to higher-derivative gravity theories with holographic $c$-theorem.

\section{Confirmations of LCFT conjecture for TMG} \label{sec:2}

The TMG action is given by \cite{Deser:1982vy,
Deser:1982sv}
\eq{
S_{\textrm{\tiny TMG}} = \frac{1}{\ka^2}\,\int \extd^3x\sqrt{|g|}\,\Big(R+\frac{2}{\ell^2}\Big) + \frac{1}{2\mu\ka^2}\,\int\extd^3x \epsilon^{\la\mu\nu}\,\Ga^\si{}_{\la\rho}\,\Big(\partial_\mu\Ga^\rho{}_{\nu\si}+\frac23\,\Ga^\rho{}_{\mu\tau}\Ga^\tau{}_{\nu\si}\Big)
}{eq:intro1}
where $\ka^2=16\pi G_N$ is the gravitational coupling, $\ell$ the AdS radius and $\mu$ the Chern--Simons coupling. 
The quantity $\epsilon$ denotes the Levi-Civita symbol.
We assume $\ell,\mu>0$ with no loss of generality and use the sign conventions of \cite{Grumiller:2008qz}.
There are two independent dimensionless combinations of the coupling constant, $\ell/G_N$ and $\mu\ell$.
Both of them enter in the values of the central charges \cite{Kraus:2005zm}.
\eq{
c_L=\frac{3\ell}{2G_N}\,\big(1-\frac{1}{\mu\ell}\big)\qquad c_R=\frac{3\ell}{2G_N}\,\big(1+\frac{1}{\mu\ell}\big)
}{eq:LCFT7}
The left central charge $c_L$ can be made vanishing by tuning. 
\eq{
\mu\ell=1
}{eq:LCFT8}
The fact that $c_L=0$ led to two
--- not necessarily contradictory --- conjectures for TMG at the critical point 
\eqref{eq:LCFT8}: the chiral gravity conjecture \cite{Li:2008dq}, according to which the dual CFT is chiral and unitary, and the LCFT conjecture \cite{Grumiller:2008qz}, according to which the dual CFT is logarithmic and non-unitary.


\subsection{Early indications}

We review now early indications 
concerning both conjectures.
To this end, consider graviton excitations $\psi$ around a global AdS$_3$ background,
\eq{
g_{\mu\nu}=g^{\textrm{\tiny AdS}}_{\mu\nu}+\psi_{\mu\nu}
}{eq:whatever}
with
\eq{
g^{\textrm{\tiny AdS}}_{\mu\nu}\,\extd x^\mu\extd x^\nu = \ell^2\,\big(\extd\rho^2-\frac14\,\cosh^2{\!\!\rho}\, (\extd u+\extd v)^2 +\frac14\,\sinh^2{\!\!\rho}\,(\extd u-\extd v)^2\big)
}{eq:cg20}
Li, Song and Strominger \cite{Li:2008dq} found an efficient way to construct them. 
Imposing transverse gauge $\nabla_\mu \psi^{\mu\nu}=0$ and defining the mutually commuting first order operators
\eq{
\big({\cal D}^M\big)_\mu^\be = \de_\mu^\be + \frac{1}{\mu}\,\varepsilon_\mu{}^{\al\be}\nabla_\al \qquad \big({\cal D}^{L/R}\big)_\mu^\be = \de_\mu^\be \pm \ell \,\varepsilon_\mu{}^{\al\be}\nabla_\al
}{eq:f22}
allows to write the linearised equations of motion around global AdS$_3$ as follows.
\eq{
({\cal D}^M{\cal D}^L{\cal D^R} \psi)_{\mu\nu} = 0
}{eq:LCFT9}
A mode annihilated by ${\cal D}^M$ (${\cal D}^L$) [${\cal D}^R$] \{$({\cal D}^L)^2$ but not by ${\cal D}^L$\} is called massive (left-moving) [right-moving] \{logarithmic\} and is denoted by $\psi^M$ ($\psi^L$) [$\psi^R$] \{$\psi^{\rm log}$\}. 
Away from the critical point, $\mu\ell\neq 1$, the general solution to the linearised equations of motion \eqref{eq:LCFT9} is obtained from linearly combining left, right and massive modes \cite{Li:2008dq}. 
At the critical point ${\cal D}^M$ degenerates with ${\cal D}^L$ and the general solution to the linearised equations of motion \eqref{eq:LCFT9} is obtained from linearly combining left, right and logarithmic modes \cite{Grumiller:2008qz}. 
The left- and right-moving modes are gauge degrees of freedom in the bulk.
The massive and logarithmic modes constitute a physical bulk degree of freedom, the massive graviton.

Interestingly, it was discovered in \cite{Grumiller:2008qz} that the modes $\psi^{\rm log}$ and $\psi^L$ behave as follows:
\eq{ 
(L_0+\bar L_0) \left(\begin{array}{c} \psi^{\rm log} \\ \psi^L
\end{array}\right) = \left(\begin{array}{c@{\quad}c}
2 & 1 \\
0 & 2
\end{array}\right) \left(\begin{array}{c} \psi^{\rm log} \\ \psi^L \end{array}\right)
}{eq:cg79a} 
where $L_0=i\partial_u$, $\bar L_0=i\partial_v$, and
\eq{ 
(L_0-\bar L_0) \left(\begin{array}{c} \psi^{\rm log} \\ \psi^L
\end{array}\right) = \left(\begin{array}{c@{\quad}c}
2 & 0 \\
0 & 2
\end{array}\right) \left(\begin{array}{c} \psi^{\rm log} \\ \psi^L \end{array}\right)
}{eq:cg1a}
With the standard definitions of the Hamiltonian $H=L_0+\bar L_0$ and the angular momentum $J=L_0-\bar L_0$ we recover exactly \eqref{eq:cg79} and \eqref{eq:cg1}. 
It was further shown that the existence of the logarithmic excitations $\psi^{\rm log}$ is not an artefact of the linearised approach, but persists in the full theory \cite{Grumiller:2008pr,Carlip:2008qh}.
Additional observations made in \cite{Grumiller:2008qz} were the degeneracy of the conformal weights of $\psi^L$ and $\psi^{\rm log}$, the finiteness of the Brown--York stress tensor, the compatibility of the logarithmic excitations with asymptotic AdS behaviour and the instability induced by the negative energy of the logarithmic excitations, which can be interpreted as a sign for non-unitary behaviour on the gravity side. 
Moreover, the Brown--York stress tensor was found to be finite, conserved, traceless \cite{Grumiller:2008qz} and non-chiral \cite{Henneaux:2009pw,Maloney:2009ck,Skenderis:2009nt,Ertl:2009ch}.
All these features were required by the purported AdS/LCFT correspondence and thus 
supported the validity of the LCFT conjecture.

An important source of inspiration for \cite{Grumiller:2008qz} was Carlip, Deser, Waldron and Wise's immediate questioning \cite{Carlip:2008jk} 
of a particular consequence of the chiral gravity conjecture: the absence of massive graviton excitations.
It turns out that the choice of boundary conditions plays a crucial role. 
For asymptotic AdS boundary conditions the chiral gravity conjecture was falsified soon \cite{Grumiller:2008qz,Grumiller:2008es,Henneaux:2009pw}:
The logarithmic mode is a propagating degree of freedom and the theory is not chiral.
However, imposing boundary conditions that are stricter than asymptotic AdS$_3$ behaviour, namely Brown--Henneaux (BH) boundary conditions \cite{Brown:1986nw}, eliminates the logarithmic mode and opens up the possibility of a chiral theory \cite{Li:2008yz,Strominger:2008dp}.
Nevertheless, Giribet, Kleban and Porrati found a specific logarithmic excitation that is compatible with BH boundary conditions, which seemed to render also this theory unstable \cite{Giribet:2008bw}.
This was, however, explained by Maloney, Song and Strominger \cite{Maloney:2009ck} to be an artefact of the linearisation: there is a linearisation instability in the theory and the linearised spectrum develops logarithmic behaviour at second order in perturbation theory, violating the BH boundary conditions. In particular this happens for the mode found by Giribet {\it et al.}

Thus the choice of boundary conditions could result in two very different theories. 
From the CFT viewpoint the restriction to BH boundary conditions can possibly be interpreted as restricting to a charge zero superselection sector \cite{Maloney:2009ck} of the LCFT dual to TMG.\footnote{The possibility to truncate the dual LCFT to a unitary subsector was already envisaged in \cite{Grumiller:2008qz}.} 
Indeed, the only contributions to the left-moving Virasoro charges come from terms with logarithmic boundary behaviour \cite{Maloney:2009ck}, so BH boundary conditions naturally restrict to this subsector. 
It is not clear whether such a subsector can be regarded as a CFT on its own \cite{Maloney:2009ck}. 
It should also be noted that there is an argument against the existence of extremal CFTs for large central charges \cite{Gaberdiel:2007ve,Gaberdiel:2008pr}. Since these are the proposed duals for chiral gravity, this may be taken
as a counter-indication against the chiral gravity conjecture.
We do not consider the truncation to chiral gravity any further in the present work (see also \cite{Andrade:2009ae,Compere:2010xu}), except for Section \ref{sec:7} where we mention new candidate theories for chiral gravity. 

From now on we instead focus on the theories obtained when allowing for arbitrary asymptotic AdS excitations.
As summarised above, many early indications pointed towards the validity of the LCFT conjecture for this case.
However, it was clear that more stringent tests of this conjecture would be desirable.
The calculation of correlators and the 1-loop partition function on the gravity side provided such tests.

\subsection{Correlators and 1-loop partition function}

If the LCFT conjecture is true then the calculation of correlators on the gravity side must be compatible with \eqref{eq:short1}.
This was indeed found to be the case \cite{Skenderis:2009nt,Grumiller:2009mw}, with the following result for the new anomaly.
\eq{
b_L = - c_R = -\frac{3\ell}{G_N}
}{eq:short2}
Also the 3-point correlators were found to be compatible with the conformal Ward identities \cite{Grumiller:2009mw}.
Thus, 2- and 3-point correlators are precisely as required for an LCFT.

\begin{figure}[t]
 \centering
 \subfigure[]{
  \includegraphics[height=6.5cm]{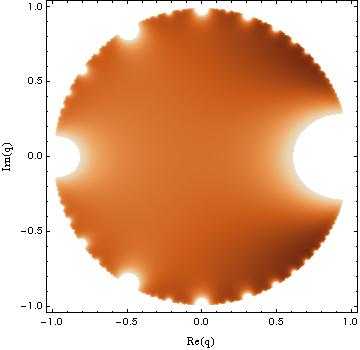}
   \label{fig:subfig1}
   }
 \subfigure[]{
  \includegraphics[height=6.5cm]{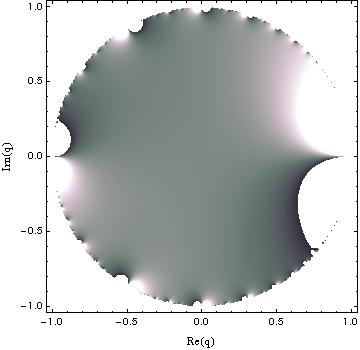}
   \label{fig:subfig2}
   }
 \caption{
  Real (a) and imaginary (b) parts of $\ln Z_{\rm TMG}^{\rm 1-loop}$. The shading goes from darker (lower values) to brighter (higher values). The plots are cut off at large positive and negative values, so the white regions along the unit circle represent poles of either sign.}  \label{fig:Z}
\end{figure}

The most recent piece of evidence came from the calculation of the 1-loop partition function in Euclidean TMG for thermal AdS \cite{Grumiller:2010xv}.
\begin{equation}
 Z_{\rm TMG}^{\rm 1-loop}(q,\bar q) = \prod\limits_{n=2}^\infty \frac{1}{|1-q^n|^2}
 \prod\limits_{m=2}^\infty\prod\limits_{\bar m=0}^\infty \frac{1}{1-q^m\bar q^{\bar m}}
 \label{eq:ZTMG}
\end{equation}
Here $q=e^{i(\tau_1+i \tau_2)}$, and the quantities $2\pi\tau_1$ and $2\pi\tau_2$ are equivalent to the angular potential $\theta$ and inverse temperature $\beta$, respectively.
Real and imaginary parts of $\ln Z_{\rm TMG}^{\rm 1-loop}$ in the unit circle are plotted in Fig.~\ref{fig:Z}.
In these density plots the self-similarity expected from a CFT partition function is clearly visible close to the boundary of the unit-circle.
The real (imaginary) part has poles (zeros) at the roots of unity.

On the LCFT side the partition function of the Virasoro descendants of the left, right and logarithmic states is \cite{Grumiller:2010xv}
\begin{equation}
Z^0_{\rm LCFT}(q,\bar q) = 
 \Bigl(1 + \frac{q^2}{|1-q|^2} \Bigr)\,\prod\limits_{n=2}^\infty \frac{1}{|1-q^n|^2} 
 \label{eq:ZLCFT}
\end{equation}
Note that the partition function $Z^0_{\rm LCFT}$ does not take into account multi-particle logarithmic excitations.
Thus, the difference between the TMG partition function \eqref{eq:ZTMG} and the contribution \eqref{eq:ZLCFT} to the full LCFT partition function must contain information about multi-graviton states.
If this interpretation is correct then the multiplicities $N_{h,\bar{h}}$ in the expression below all must be non-negative.
\begin{equation}
Z_{\rm TMG} = \prod\limits_{n=2}^\infty \frac{1}{|1-q^n|^2}\, 
\prod\limits_{m=2}^\infty\prod\limits_{\bar m=0}^\infty \frac{1}{1-q^m\bar q^{\bar m}} 
 =  Z^0_{\rm LCFT}
+ \sum_{h,\bar{h}=0}^\infty N_{h,\bar{h}}\, q^h \bar{q}^{\bar{h}} \, \prod_{n=1}^{\infty} \frac{1}{|1-q^n|^2} 
\label{eq:angelinajolie}
\end{equation}
This was indeed found to be the case \cite{Grumiller:2010xv} and provides a fairly non-trivial check on the validity of the LCFT conjecture.\footnote{Note that in the computations of \cite{Grumiller:2010xv}, the Hilbert space trace is taken over all normalisable modes, including the logarithmic excitations. The fact that the partition function is consistent with an LCFT interpretation should therefore not be considered as a counter-argument against the proposal that the theory is chiral if BH boundary conditions are imposed.} 

Collecting all evidence so far it is fair to state that the LCFT conjecture is likely to be correct for TMG \eqref{eq:intro1} at the critical point \eqref{eq:LCFT8}.

\section{LCFT conjecture for other massive gravity theories} \label{sec:3}

To be able to put AdS/LCFT to practical use, it is of course necessary to have several examples of the correspondence available.
Note furthermore that the important class of applications where the stress-energy tensor acquires a logarithmic partner can be addressed only if a metric mode degenerates.
Thus, it is of prime importance to study higher curvature theories and look for LCFT behaviour.
Below we expound the status for two popular higher curvature theories in three dimensions: NMG and GMG.

\subsection{New massive gravity}

NMG \cite{Bergshoeff:2009hq} is a 3-dimensional gravitational theory with massive spin-2 excitations.
It exhibits similar features to TMG.
The main difference to TMG is that NMG does not violate parity.
Otherwise, as we shall recapitulate below, the story of NMG is very similar to the TMG story.
In particular, there is again striking evidence for a dual LCFT, provided the parameters in the action are tuned to a critical value.
We summarise now this evidence.

The action for NMG is given by \cite{Bergshoeff:2009hq} 
\eq{
S_{\textrm{\tiny NMG}}=\frac{1}{\ka^2}\,\int\extd^3x\sqrt{-g}\,\Big[\si R+\frac{1}{m^2}\,\big(R^{\mu\nu}R_{\mu\nu}-\frac38\,R^2\big)-2\la m^2\Big]
}{eq:NMG1} 
where $m$ is a mass parameter, $\lambda$ a dimensionless cosmological parameter and $\sigma=\pm 1$ the sign of the Einstein--Hilbert term. 
This action leads to equations of motion that have as particular solutions AdS$_3$ \eqref{eq:cg20}. 
The AdS radius is given by $1/\ell^2=2m^2(\si\pm\sqrt{1+\la})$.
For positive $\la$ there is always a unique AdS$_3$ vacuum.
We assume henceforth $\la>0$, unless stated otherwise.
There are two independent dimensionless combinations of the coupling constants, $\ell/G_N$ and $m^2\ell^2$. 
Both of them enter in the values of the central charges \cite{Liu:2009bk,Bergshoeff:2009aq}
\eq{
c_L \ = \ c_R \ =\ \frac{3\ell}{2G_N}\,\Big(\sigma+\frac{1}{2m^2\ell^2}\Big)
}{eq:NMG6}
Both central charges can be made vanishing by tuning. 
\eq{
 2m^2\ell^2 = -\si
}{eq:NMG4}
Given our experience with TMG, this provides a first hint that for NMG the dual CFT might be logarithmic at the critical point \eqref{eq:NMG4}. 
The linearised equations of motion for NMG around AdS$_3$ are similar to the linearised equations of motion for TMG \eqref{eq:LCFT9} \cite{Liu:2009bk}
\eq{
({\cal D}^L{\cal D}^R{\cal D}^M{\cal D}^{\tilde M}\psi)_{\mu\nu}=0
}{eq:NMG2}
with the mutually commuting first order operators
\eq{
\big({\cal D}^{M/\tilde M}\big)_\mu{}^\be = \de_\mu{}^\be \pm \frac{1}{M}\,\varepsilon_\mu{}^{\al\be}\nabla_\al \qquad \big({\cal D}^{L/R}\big)_\mu{}^\be = \de_\mu{}^\be \pm \ell \,\varepsilon_\mu{}^{\al\be}\nabla_\al 
}{eq:NMG3}
where $M$ is determined from the parameters in the action as
\eq{
M\ell = \sqrt{\frac{1}{2} - \si m^2 \ell^2}
}{eq:NMG3.5}
If they are tuned to the critical point \eqref{eq:NMG4} then $M \ell = 1$. 
Consequently, the operators ${\cal D}^M$ and ${\cal D}^L$ degenerate, and analogously do ${\cal D}^{\tilde M}$ and ${\cal D}^R$. This degeneration of operators provides another hint that the dual theory might be an LCFT.
Furthermore, the consistency of logarithmic boundary conditions has been demonstrated by Liu and Sun \cite{Liu:2009kc}.
Note that there is another point of degeneracy: If $2m^2\ell^2 = \si$ then $M$ vanishes and the two massive modes degenerate. Because of the many special features of this theory we shall study it separately in Section \ref{sec:6}.

The 2-point correlators were calculated on the gravity side at the critical point \eqref{eq:NMG4} in \cite{Grumiller:2009sn,Alishahiha:2010bw}.
The non-vanishing ones are given by
\begin{align}
& \langle\psi^{\rm log}(z,\bar z)\,\psi^{L}(0)\rangle \ = \ \frac{b_L}{2z^4} \\
& \langle\psi^{\widetilde{\rm log}}(z,\bar z)\,\psi^{R}(0)\rangle \ = \ \frac{b_R}{2\bar z^4} \\
& \langle\psi^{\rm log}(z,\bar z)\,\psi^{\rm log}(0,0)\rangle \ = \ -\frac{b_L\ln{(m_L^2|z|^2)}}{z^4} \\
& \langle\psi^{\widetilde{\rm log}}(z,\bar z)\,\psi^{\widetilde{\rm log}}(0,0)\rangle \ = \ -\frac{b_R\ln{(m_R^2|z|^2)}}{\bar z^4}
\end{align}
with the new anomalies
\eq{
b_L \ = \ b_R \ = \ -\si\,\frac{12\ell}{G_N}
}{eq:NMG14}
These are precisely the correlators of a parity invariant LCFT.

The 1-loop partition function was calculated in \cite{Grumiller:2010xv}.
\begin{equation}
Z_{\textrm{NMG}}(q)= \prod\limits_{n=2}^\infty \frac{1}{|1-q^n|^2}
\prod\limits_{m=2}^\infty\prod\limits_{\bar m=0}^\infty \frac{1}{1-q^m\bar q^{\bar m}} \prod\limits_{l=0}^\infty\prod\limits_{\bar l=2}^\infty \frac{1}{1-q^l\bar q^{\bar l}} 
\label{eq:ZNMGfinal}
\end{equation}
The result \eqref{eq:ZNMGfinal} can now be compared with the partition function of the LCFT dual, analogue to the discussion for TMG above. 
Again all multiplicity coefficients $N_{h,\bar h}$ in the expression analogue to \eqref{eq:angelinajolie} turned out to be positive. 

Collecting all evidence so far it is fair to state that the LCFT conjecture is likely to be correct for NMG \eqref{eq:NMG1} at the critical point \eqref{eq:NMG4}.

\subsection{Generalised massive gravity}

GMG is NMG plus the Chern--Simons term of TMG.
The GMG action reads \cite{Bergshoeff:2009hq} 
\eq{
S_{\textrm{\tiny GMG}}=S_{\textrm{\tiny NMG}}+\frac{1}{2\mu\ka^2}\,\int\extd^3x \epsilon^{\la\mu\nu}\,\Ga^\si{}_{\la\rho}\,\Big(\partial_\mu\Ga^\rho{}_{\nu\si}+\frac23\,\Ga^\rho{}_{\mu\tau}\Ga^\tau{}_{\nu\si}\Big)
}{eq:GMG1} 
Again, AdS$_3$ \eqref{eq:cg20} arises as a solution of the classical equations of motion. 
The AdS radius again is given by $1/\ell^2=2m^2(\si\pm\sqrt{1+\la})$.
Again we assume $\la>0$ in order to have a unique AdS$_3$ vacuum.
A relevant difference to previous cases is that now there are three independent dimensionless combinations of the coupling constants, $\ell/G_N$, $m^2\ell^2$ and $\mu\ell$.
All of them enter in the values of the central charges \cite{Liu:2009pha,Bergshoeff:2009aq}.
\eq{
c_L = \frac{3\ell}{2G_N}\,\Big(\si+\frac{1}{2m^2\ell^2}-\frac{1}{\mu\ell}\Big) \qquad c_R = \frac{3\ell}{2G_N}\,\Big(\sigma+\frac{1}{2m^2\ell^2}+\frac{1}{\mu\ell}\Big)
}{eq:GMG6}
The left central charge $c_L$ can be made to vanish by tuning.
\eq{
\frac{1}{\mu\ell}-\frac{1}{2m^2\ell^2} = \si
}{eq:GMG4}
Given our experiences with TMG and NMG it seems reasonable to conjecture that for GMG at the critical line \eqref{eq:GMG4} the dual CFT is an LCFT.
The linearised equations of motion for GMG around AdS$_3$ are similar to the linearised equations of motion for NMG \eqref{eq:NMG2} \cite{Bergshoeff:2009aq}
\eq{
({\cal D}^L{\cal D}^R{\cal D}^{m_1}{\cal D}^{m_2}\psi)_{\mu\nu}=0
}{eq:GMG2}
with the mutually commuting first order operators ${\cal D}^{L/R}$ as before \eqref{eq:NMG3}, and
\eq{
\big({\cal D}^{m_1}\big)_\mu{}^\be = \de_\mu{}^\be + \frac{1}{m_1}\,\varepsilon_\mu{}^{\al\be}\nabla_\al \qquad \big({\cal D}^{m_2}\big)_\mu{}^\be = \de_\mu{}^\be + \frac{1}{m_2}\,\varepsilon_\mu{}^{\al\be}\nabla_\al 
}{eq:GMG3}
where $m_1, m_2$ are determined from the parameters in the action:
\eq{
m_{1,2}\ell = \frac{m^2 \ell^2}{2\mu\ell} \pm \sqrt{\frac{1}{2} - \si m^2 \ell^2 + \frac{m^4\ell^4}{4 \mu^2 \ell^2}}
}{eq:GMG3.5}
If they are tuned to the critical line \eqref{eq:GMG4} then $m_1\ell=1$.
Consequently, the operators ${\cal D}^{m_1}$ and ${\cal D}^L$ degenerate, as expected for an LCFT.
In GMG there is a whole line where the operators ${\cal D}^{m_1}$ and ${\cal D}^{m_2}$ degenerate:
\eq{
\frac{m^2\ell^2}{4\mu^2\ell^2} + \frac{1}{2m^2\ell^2} = \si
}{eq:GMG5}
This has two interesting consequences. First, as in NMG this kind of degeneration allows for the possibility of an LCFT with non-vanishing central charges. Here however, the theory is in general not parity invariant: $c_L \neq c_R$. 
Second, at the intersection of the critical lines \eqref{eq:GMG4} and \eqref{eq:GMG5} there is a critical point
\eq{
m^2\ell^2=2\mu\ell=\frac32\,\si
}{eq:GMG7}
where three operators degenerate, ${\cal D}^{m_1}$, ${\cal D}^{m_2}$ and  ${\cal D}^L$, and the left central charge vanishes, $c_L=0$.
If the LCFT conjecture is correct this should lead to a rank 3 Jordan cell generalising the rank 2 structure \eqref{eq:cg79}.
Consequently, there should be two partners for the holomorphic part of the energy-momentum tensor ${\cal O}^L$, a logarithmic mode ${\cal O}^{\rm log}$ and a square-logarithmic mode $\Osq$.
On the gravity side this should be reflected in boundary conditions that are even more relaxed than the logarithmic boundary conditions of \cite{Grumiller:2008es}.
Indeed, such boundary conditions were found (and their consistency was shown) by Liu and Sun \cite{Liu:2009pha}. 

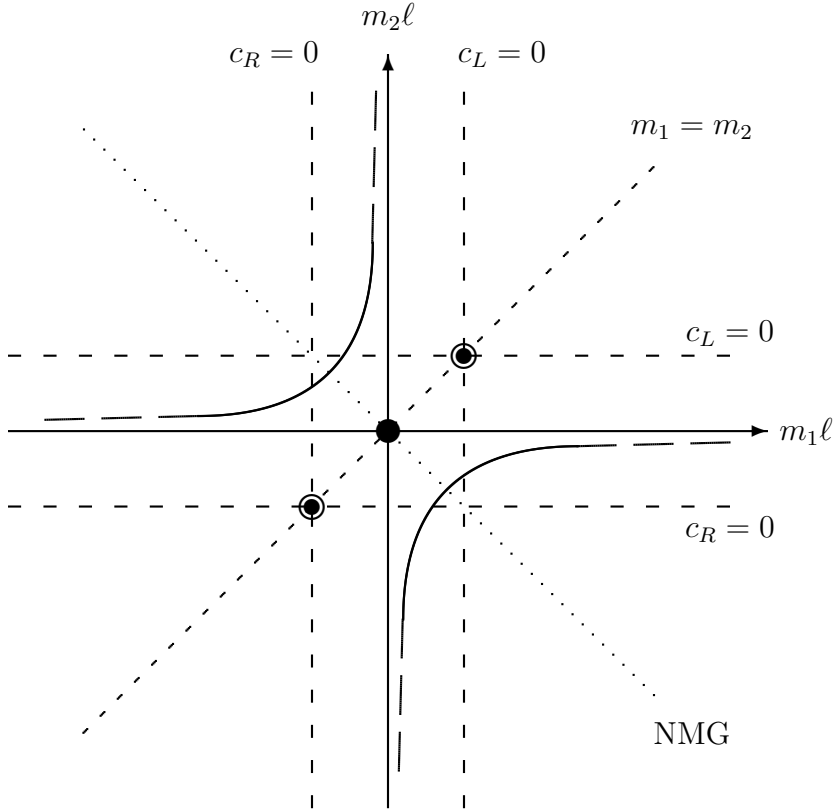
\begin{figure}
\begin{center}
\setlength{\unitlength}{1mm}
\begin{picture}(100,100)(-10,-10)
\thicklines
\put(40,-10){\vector(0,1){100}} 
\put(40,95){\makebox(0,0){$m_2\ell$}}

\put(-10,40){\vector(1,0){100}} 
\put(95,40){\makebox(0,0){$m_1\ell$}}

\dashline{2}(-10,50)(85,50)
\put(85,53){\makebox(0,0){$c_L =0 $}}

\dashline{2}(50,-10)(50,85)
\put(55,90){\makebox(0,0){$c_L =0 $}}

\dashline{2}(-10,30)(85,30)
\put(25,90){\makebox(0,0){$c_R =0 $}}

\dashline{2}(30,-10)(30,85)
\put(85,27){\makebox(0,0){$c_R =0 $}}

\dashline{1}(0, 0)(75,75)
\put(80,80){\makebox(0,0){$m_1 = m_2$}}

\dottedline{2}(0, 80)(75 ,5)
\put(80,0){\makebox(0,0){NMG}}

\put(40,40){\circle*{3}}
\put(50,50){\circle{3}}
\put(50,50){\circle*{2}}

\put(30,30){\circle{3}}
\put(30,30){\circle*{2}}
 
\qbezier(42,15)(42,38)(65,38)
\qbezier(38,65)(38,42)(15,42)
\dashline{5}(65,38)(85, 38.5)
\dashline{5}(42,15)(41.5, -5)
\dashline{5}(38,65)(38.5, 85)
\dashline{5}(15,42)(-5, 41.5)

\end{picture}

\caption{Special loci in the GMG parameter space. See main text for explanations. }

 \label{fig:2}
\end{center}

\end{figure}

For future reference it is convenient to display the important loci in the GMG parameter space described above. Figure \ref{fig:2} provides such a plot. For convenience we choose the parameters to be $m_1$ and $m_2$ of \eqref{eq:GMG3.5}.
The theory reduces to NMG along $m_1 = -m_2$ (dotted). Along the four lines $m_{1,2}\ell = \pm 1$ (dashed) a massive mode degenerates with a
boundary graviton and becomes zero norm. Here, the stress-energy tensor acquires a logarithmic partner. Along the line $m_1 = m_2$
(dashed) the two massive modes degenerate resulting in LCFTs where the central charges are, in general, non-zero and unequal. At the special
points $m_1\ell = m_2 \ell = \pm 1$ (double circles) three modes degenerate, and the LCFT has a rank three Jordan cell. 
The two hyperbolas represent the loci where $m_1 m_2 \ell^2 = -1/2$, related to the singular limit $m^2 \to 0$.
Finally, at the origin (circle) the theory is logarithmic, parity invariant, partially massless and
has non-zero central charges. We call this theory ``partially massless gravity'' and study it in some detail in Section \ref{sec:6}.

Collecting all evidence so far there is reasonable evidence for the LCFT conjecture in GMG. Interestingly, there are now three qualitatively different possibilities for the dual LCFT: if we are on the critical line \eqref{eq:GMG4} we expect a ``standard'' dual LCFT similar to the one arising in TMG with a rank 2 Jordan cell and a logarithmic partner for the holomorphic part of the energy-momentum tensor. If we are on the critical line \eqref{eq:GMG5} we expect an ``exotic'' dual LCFT where the massive operators degenerate with each other, but not with the energy-momentum tensor. If we are on the critical point \eqref{eq:GMG7} we expect a ``standard'' dual LCFT, but with a rank 3 Jordan cell. We shall exhibit all these features in detail while calculating the new anomalies for GMG. Before doing this we describe a convenient short-cut to determine the new anomalies.

\section{Short-cut to new anomalies} \label{sec:4}

We discuss now a short-cut to determine the new anomalies for LCFTs that arise as limits of parameter families of ordinary CFTs. 
We first describe the general procedure and then apply it to two examples.
The limiting constructions below are themselves not new, see e.g.~\cite{Rasmussen:2004gx, Rasmussen:2004na}. In this section we limit ourselves to the case where two operators degenerate, postponing the more general case until section \ref{sec:5}. 

\subsection{Introducing the short-cut}

We start by considering the CFT side (see also \cite{Grumiller:2010rm}). Consider a family\footnote{Note that by Zamolodchikov's $c$-theorem \cite{Zamolodchikov:1986gt}, the central charge is always constant in a continuous family of unitary CFTs. The models of interest here are, however, not unitary, and the central charges vary continuously. LCFTs can also arise as limits of unitary CFTs if there is an accumulation point in parameter space.} of CFTs that depends on some (possibly multidimensional) continuous parameter $\vecm$. 
Suppose we have two operators ${\cal O}^{i}(\vecm)$ $i=1,2$ with different conformal weights $h^i(\vecm),\,\bar h^i(\vecm)$.
The non-vanishing 2-point correlators are then given by
\eq{
\langle{\cal O}^{i}(z,\bar z){\cal O}^{i}(0,0)\rangle = \frac{c_i(\vecm)}{2z^{2h^i(\vecm)}\bar z^{2\bar h^i(\vecm)}}
}{eq:short12}
with some normalisation constants $c_i(\vecm)$. Assume now that there is a point $\vecm_0$ where these two operators 
degenerate: $\cO^1(\vecm_0) = \cO^2(\vecm_0)$. In order to perform a limit construction we choose some path in parameter space $\vecm(\epsilon)$ parametrised by $\epsilon \in \mathbb{R}$ such that $\vecm(0) = \vecm_0$. Without loss of generality we may choose $\epsilon$ to be simply related to the difference between the weights:
\eq{
h^1\big(\vecm(\epsilon)\big) - h^2\big(\vecm(\epsilon)\big) = \bar{h}^1\big(\vecm(\epsilon)\big) - \bar{h}^2\big(\vecm(\epsilon)\big) \equiv \De_{12} = \epsilon
}{eq:short12.5}
Note that the first equality follows from requiring locality, implying that $h - \bar{h} \in \mathbb{Z}$.
Essential for the appearance of a logarithmic pair at this point is that the constants $c_i$ vanish linearly 
in $\epsilon$ with related slopes:
\eq{
c_1 = B \epsilon + \ldots \qquad c_2 = -B \epsilon + \ldots
}{eq:short12.6}
It is convenient, and always possible, to rescale the operators $\cO^i$ by functions $f^i(\epsilon)$ in such a way that the ellipses in the above equations actually vanish identically. (Nonvanishing ellipses correspond to adding a multiple of $\cO^1$ to $\cO^{\rm log}$, see below.) Note, however, that we must keep $f^i(0) = 1$. This ensures the coincidence of the operators $\cO^i$ with each other at $\vecm_0$ and guarantees that their predefined normalisation is maintained, which is needed to have a well-defined new anomaly.

Let us now introduce a logarithmic operator.
\eq{
{\cal O}^{\rm log} := \lim_{\De_{12}\to 0} \frac{{\cal O}^{1}-{\cal O}^{2}}{\De_{12}}
}{eq:short10} 
With the above assumptions it is straightforward to obtain the following non-vanishing 2-point correlators at $\vecm_0$
\begin{subequations}
\label{eq:short8}
\begin{align}
& \langle{\cal O}^{\rm log}(z,\bar z){\cal O}^{1}(0,0)\rangle = \frac{B}{2z^{2h(\vecm_0)}\bar z^{2\bar h(\vecm_0)}} \\
& \langle{\cal O}^{\rm log}(z,\bar z){\cal O}^{\rm log}(0,0)\rangle = -\frac{B \ln{(m_L|z|^2)}}{z^{2h(\vecm_0)}\bar z^{2\bar h(\vecm_0)}}
\end{align}
\end{subequations}
For $h=2, \bar h=0$ we recover precisely the LCFT correlators \eqref{eq:Llog} and \eqref{eq:loglog}; 
for different values of the weights we get the correct generalisations of these correlators.
The new anomaly is then determined by  
\eq{
b_L = B = \lim_{\De_{12} \to 0} \frac{c_1}{\De_{12}} = \lim_{\De_{21} \to 0} \frac{c_2}{\De_{21}}
}{eq:short13}
In summary, the new anomaly is determined by the ratio between the constants $c_i$ and the difference $\De_{12}$ between the weights in the critical limit. Note, finally, that the fact that there are only two linearly independent operators at $\vecm_0$ immediately implies that any parametrisation $\vecm(\epsilon)$ must give the same result.

If one of the operators (say, $\cO^1$) is a flux component of the energy-momentum tensor the information required for determining the new anomaly is easily accessible on the gravity side. In this case the preferred normalisation is one in which $c_1$ is the central charge, a quantity that can be derived in a number of ways. For the gravity models considered here, TMG, NMG and GMG, the results are well-known. 
Moreover, all left-, right-moving and massive modes are solutions of the first order partial differential equation
\eq{
({\cal D}^{m_i} \psi)_{\mu\nu} = \psi_{\mu\nu} + \frac{1}{m_i}\,\,\varepsilon_\mu{}^{\al\be}\nabla_\al \psi_{\be\nu}= 0
}{eq:short14}
The conformal weights of normalisable primaries $\psi$ that solve \eqref{eq:short14} were calculated in \cite{Li:2008dq}.
\begin{subequations}
\label{eq:short15}
\begin{align}
m_i>0:\quad \big(h,\bar h\big) = \Big(\frac{3+ m_i\ell}{2},\, \frac{-1+ m_i\ell}{2}\Big) \\
m_i<0:\quad \big(h,\bar h\big) = \Big(\frac{-1- m_i\ell}{2},\,\frac{3- m_i\ell}{2}\Big)
\end{align}
\end{subequations}
For $m_1\ell=1$ ($m_1\ell=-1$) we recover the weights of the normalisable left-moving (right-moving) primaries.
In this case inserting the result for the weights \eqref{eq:short15} into the definition \eqref{eq:short12.5} one obtains 
\eq{
\De_{12} = \frac{1}{2} \, \big(1-|m_2\ell|\big)
}{eq:short20}
We now illustrate the short-cut by applying it to two known examples.

\subsection{Known examples}

Let us first apply the short-cut described above to TMG. 

In that case $m_1=1/\ell$ and $m_2=\mu$, and thus $\De_{12} = (1-\mu\ell)/2$ (we take $\mu > 0$).
From \eqref{eq:LCFT7} and \eqref{eq:short13} we obtain
\eq{
b_L = \lim_{\mu\ell\to 1}\frac{3\ell}{2G_N} \,\frac{2\left(1-\frac{1}{\mu\ell}\right)}{1-\mu\ell}= -\frac{3\ell}{G_N}
}{eq:short16}
Thus, we recover indeed the result for the new anomaly \eqref{eq:short2} that was derived by calculating 2-point correlators on the gravity side.

Second, we consider NMG.
In that case $m_1=1/\ell$ and $m_2=M$ [see \eqref{eq:NMG3.5}], and thus $2 \De_{12} = 1-\sqrt{1/2-\si m^2\ell^2}$.
From \eqref{eq:NMG6} and \eqref{eq:short13} we obtain
\eq{
b_L = \lim_{m_2 \ell\to 1} \, \frac{3\ell}{G_N}\,\frac{\si+\frac{1}{2m^2\ell^2}}{1-\sqrt{1/2-\si m^2\ell^2}} = -\si\,\frac{12\ell}{G_N}
}{eq:short18}
Thus, we recover indeed the result for the new anomalies \eqref{eq:NMG14} that was derived by calculating 2-point correlators on the gravity side.

\section{New anomalies in generalised massive gravity} \label{sec:5}

The short-cut described in the previous section led to the correct values for the new anomalies in TMG and NMG.
In this section we apply this short-cut to GMG, where no results for new anomalies exist so far.
We stress that the short-cut by no means can be used as evidence in favour of the LCFT conjecture.
Rather, in this section we {\em assume} that there are LCFT duals for GMG with appropriate tuning of the coupling constant, and we merely employ the short-cut to determine the new anomalies of the putative LCFT duals.

For GMG there are three qualitatively different kinds of LCFTs that can occur: A single massive mode can degenerate with a boundary graviton paralleling the TMG case; both massive modes can degenerate with a boundary graviton yielding a rank three Jordan cell; and two massive modes can degenerate with each other leading to an LCFT where the energy-momentum tensor does not have a logarithmic partner. We treat these cases in turn.
 
Let us begin by recording the central charges \eqref{eq:GMG6} expressed in the masses $m_{1,2}$ in \eqref{eq:GMG3.5}:
\eq{
c_{L} = \frac{3\ell \si}{G_N}\frac{(1 - m_1\ell)(1 - m_2\ell)}{1 + 2m_1 m_2 \ell^2} \qquad c_{R} = \frac{3\ell \si}{G_N}\frac{(1 + m_1\ell)(1 + m_2\ell)}{1 + 2m_1 m_2 \ell^2}
}{eq:news1}
From these expressions it is clear that the new anomaly is non-zero along the lines $m_{1,2}\ell = 1$ except at the doubly critical point $m_1\ell = m_2\ell = 1$. In the language of Section \ref{sec:4} the coefficient in the correlator vanishes as $\sim \epsilon^2$ here and we need to go to second order in the expansion to extract the non-trivial quantity. We shall perform this construction in Subsection \ref{ssec:rank3} below.

\subsection{Rank 2 standard case: $m_1  \ell = 1$}

For definiteness let us consider the critical line $m_1\ell = 1$ ($m_2\ell=1$ works in the same way, with $1\leftrightarrow 2$ and $L\leftrightarrow R$). Then $c_L = 0$ and $\psi^{m_1}$ degenerates with $\psi^L$. We then have $\De_{12} = (1-m_1\ell)/2$ and thus
\eq{
b_L = \lim_{m_1\ell \to 1} \frac{6 \ell \si}{G_N}\frac{1-m_2\ell}{1+2 m_1 m_2 \ell^2} = 
\frac{6 \ell \si}{G_N}\frac{1-m_2\ell}{1+2 m_2 \ell} = 
\frac{3 \ell }{G_N}\left( \frac{3}{\mu \ell}-4\si \right)
}{eq:news2}
This is our conjectured value for the new anomaly along the line $m_1 \ell = 1$. 
As consistency checks we recover from the GMG new anomaly \eqref{eq:news2} the NMG result \eqref{eq:short18} in the limit $\mu\to\infty$ and the TMG result \eqref{eq:short16} in the limit $m_2\to\infty$ (with $\si=+1$).

\subsection{Rank 3 standard case: $m_1  \ell = m_2 \ell = 1$}
\label{ssec:rank3}

Now we consider the case when three operators degenerate. This is realized at the doubly degenerate point \eqref{eq:GMG7} of GMG. 
First we keep the discussion quite general and apply our result to GMG at the end.

Consider again a family of CFTs parametrised by $\vecm$ and assume that three operators $\cO^i$, $i=1,2,3$, degenerate at $\vecm_0$, with 2-point correlators as in \eqref{eq:short12}. Consider a two-dimensional surface in $\vecm$-space containing $\vecm_0$ and assume that in this surface there are three curves, all meeting at $\vecm_0$, along which the pairs $(\cO^1, \cO^2)$, $(\cO^1, \cO^3)$ and $(\cO^2, \cO^3)$ degenerate, respectively. For a visualisation see Fig.~\ref{fig:2} describing the GMG situation and let, e.g., $\cO^{1,2} = \cO^{m_{1,2}}$ and $\cO^3 = \cO^L$.
The limit to $\vecm_0$ corresponds to the upper-right double circle in that picture.

Again it is convenient to parametrise the surface by parameters $\epsilon_{1,2}$ simply related to the difference in conformal weights between the operators:
\eq{
\De_{12} = \epsilon_1 \qquad \De_{23} = \epsilon_2 \qquad \De_{13} = \epsilon_1 + \epsilon_2
}{eq:three1}
with $\De_{ij} \equiv h^i - h^j$. 
We assume that at the critical curves the corresponding $c_i$ vanish. 
This implies that the $c_i$ vanish quadratically at $\vecm_0$:
\eq{
c_1 = B_1 \De_{12}\De_{13} + \ldots \qquad c_2 = B_2 \De_{21}\De_{23} + \ldots \qquad c_3 = B_3 \De_{31}\De_{32} + \ldots
}{eq:three2}
for some constants $B_i$. Again, suitable rescalings eliminate the ellipses, but should not be used to change the $B_i$. The non-trivial condition required for well-defined correlators is 
\eq{
B_1 = B_2 = B_3 \equiv B
}{eq:three3}
which is implied by imposing \eqref{eq:short12.6} along each of the critical lines.
Under these assumptions, defining operators based on limit definitions of the first and second derivatives
\eq{
\cO^{\rm log} = \lim_{\De_{12}\to 0} \, \frac{\cO^{m_1}-\cO^{m_2}}{\De_{12}} \qquad
 \Osq = \lim_{\epsilon_{1,2}\to 0} \, \frac{\De_{32} \cO^{m_1} + \De_{13} \cO^{m_2}+ \De_{21} \cO^{m_3}}{\De_{12}\De_{13}\De_{32}}\\
}{eq:three4}
gives sensible results.  
The operators $\cO^{\rm log}$ and $\Osq$, together with e.g.~$\cO^1$, constitute a basis for the operators at $\vecm_0$. 
Their 2-point correlators work out to
\begin{subequations}
\label{eq:three5}
\begin{align}
& \langle \cO^1 (z,\bar z)\cO^{\rm log}(0,0)\rangle = 0 \\
& \langle \cO^1 (z,\bar z)\Osq(0,0) \rangle =
\langle \cO^{\rm log} (z,\bar z)\cO^{\rm log}(0,0) \rangle = \frac{B}{2z^{2h(\vecm_0)}\bar{z}^{2\bar{h}(\vecm_0)}} \\
&\langle \cO^{\rm log} (z,\bar z)\Osq(0,0) \rangle = -\frac{B \log |z|^2 }{z^{2h(\vecm_0)}\bar{z}^{2\bar{h}(\vecm_0)}} \\
&\langle \Osq (z,\bar z)\Osq(0,0) \rangle = \frac{B \log^2 |z|^2 }{z^{2h(\vecm_0)}\bar{z}^{2\bar{h}(\vecm_0)}}  \label{eq:log2corr}
\end{align}
\end{subequations}
This is precisely the expected form of the correlators in an LCFT with a rank three Jordan cell. (See e.g.~\cite{RahimiTabar:1996ub,Khorrami:1997ci}.) 
The quantity $B$ generalises the new anomaly to the rank three case and is well-defined only after a normalisation has been chosen for the operators. 
It is then obtained as a limit.
\eq{
B = \lim_{\epsilon_{1,2}\to 0} \frac{c_1}{\De_{12}\De_{13}} = \lim_{\epsilon_{1,2}\to 0} \frac{c_2}{\De_{21}\De_{23}} 
= \lim_{\epsilon_{1,2}\to 0} \frac{c_3}{\De_{31}\De_{32}}
}{eq:three6} 
Let us apply this formula to the doubly degenerate point of GMG. As before a preferred normalisation is provided by requiring that the 2-point correlator of two left modes be proportional to the central charge $c_L$.
Choosing $\cO^{1,2} \sim \psi^{m_{1,2}}$ and $\cO^3 \sim \psi^L$ we have 
\eq{
\De_{31} = (1 - m_1 \ell)/2 \qquad \De_{32} = (1 - m_2 \ell)/2 \qquad c_{3} = 
\frac{3\ell \si}{G_N}\frac{(1 - m_1 \ell)(1 - m_2 \ell)}{1 + 2m_1m_2\ell^2}  
}{eq:three7}
and, consequently,
\eq{
B = \lim_{m_{1,2}\ell\to 1} \frac{12\ell \si}{G_N}\frac{1}{1 + 2m_1m_2\ell^2} = \frac{4\ell \si}{G_N}
}{eq:three8} 
Obviously it would be interesting to compute the correlators using the full machinery of AdS/CFT to confirm these results. While we do not engage in such a computation here, we present in appendix \ref{app:double} a first step in this direction: the asymptotic expansion for the doubly logarithmic mode $\psi^{\rm log^2}$ for arbitrary weights $(h,\bar{h})$.

Let us now turn to the case when two massive modes degenerate, but are distinct from any of the boundary gravitons.

\subsection{Rank 2 exotic case: $m_1  \ell = m_2 \ell \neq 1$}
As already explained, the cases in which the short-cut works best is when one of the degenerating operators is a flux component of the energy-momentum tensor. Then there is a preferred normalisation, and the correlators are determined by the central charge, which is easily accessible. When two massive modes degenerate this is no longer the case, and one must, for a given normalisation, compute the 2-point correlator between two massive modes in the vicinity of the critical locus. The new anomaly for this normalisation is then obtained as in \eqref{eq:short13}. 

It is of considerable interest to study also this situation, since the corresponding theories represent putative gravity duals of LCFTs with non-zero central charges. For GMG the two massive modes degenerate along the line $m_1 = m_2$. Had the 2-point correlators been known, \eqref{eq:short13} would have given immediately the new anomaly.

Since computing 2-point correlators is quite technical \cite{Skenderis:2009nt, Alishahiha:2010bw} when tricks as those used in \cite{Grumiller:2009mw, Grumiller:2009sn} are unavailable, we shall blithely permit ourselves another short-cut. As explained in \cite{Skenderis:2009nt} the 2-point correlator of two massive modes is related by Ward identities to the energy of the corresponding mode. The energy is also obtained from the quadratic action put on-shell, but with the important simplification that no boundary terms need be considered. This is so because the modes in question are normalisable. The only downside of this procedure is that it is not clear how to relate the normalisation of the normalisable modes to that of the non-normalisable ones used in the honest-to-God computation. 

In a higher derivative theory as GMG the Hamiltonian is conveniently obtained through the Ostrogradsky procedure as described in \cite{Li:2008dq}. Our starting point is the quadratic action (to reduce clutter we set $16\pi G_N = 1$ in this subsection unless explicit) for fluctuations $h_{\mu\nu}$ around the AdS background.
\eq{
S_{(2)} = -\frac{m_1 m_2 }{2m^2 \ell^2}\int \extd^3x\sqrt{-g}\, h^{\mu\nu} (\cD^{m_1} \cD^{m_2} \cD^{L} \cD^{R} h)_{\mu\nu}
}{eq:mass1}
By analogy to \cite{Li:2008dq} we now perform the Legendre transformation, drop certain boundary terms and get the Hamiltonian
\eq{
H = \int \extd^2x \sqrt{-g}\, \cH(h_{\mu\nu}),
}{eq:mass2}
which, evaluated on the modes in question gives the corresponding energies. 
One arrives at (a dot denotes time derivative) \cite{Zojer:thesis}
\eq{
E_{m_1} =  \si \, \frac{(1-m_1\ell)(1 + m_1\ell)(m_2 - m_1)}{1 + 2m_1m_2 \ell^2} \int \extd^2 x \sqrt{-g} \, \dot{h}^1_{\mu\nu}\eps^{\mu 0}{}_{\be}h^{\be\nu}_{1}
}{eq:mass3}
The result for $E_{m_2}$ is obtained by replacing $1 \leftrightarrow 2$ everywhere in the above equation.
We have checked that in GMG there is no choice of parameters for which all central charges and energies $E_{m_1}, E_{m_2}$ simultaneously are strictly positive.
This is the result anticipated from TMG and NMG.
The integral in \eqref{eq:mass3} is positive definite and thus does not change the structure of zeros or the sign 
of the energies entailed in the prefactors of the integral in \eqref{eq:mass3}.

Therefore, in some suitable normalisation, the constant in the 2-point correlators is
\eq{
c_{1} = \frac{3\ell \si}{2G_N} \frac{(1-m_1\ell)(1 + m_1\ell)(m_2\ell - m_1\ell)}{1 + 2m_1m_2 \ell^2}
}{eq:mass5}
and similarly with $1 \leftrightarrow 2$ for $c_2$. 
We see that the $c_i$ fulfil all expected properties: 
When the mode $h_i$ degenerates with another mode $c_i$ vanishes. 
Moreover \eqref{eq:short12.6} holds along the line $m_1 = m_2$.
We are thus in a position to compute the new anomaly along this line. 
Since $2\De_{12} = |m_1|\ell - |m_2|\ell$ we get
\eq{
b = -{\rm sign}(m_1) \frac{3\ell \si}{G_N} \frac{(1-m_1\ell)(1 + m_1\ell)}{1 + 2m_1^2 \ell^2}
}{eq:mass6}
where 
\eq{
m_1\ell = m_2\ell = \frac{m^2\ell^2}{2\mu\ell} = \pm \sqrt{\si m^2 \ell^2 - 1/2}
}{eq:mass7}
Note that the awkward looking factor ${\rm sign}(m_1)$ comes from the fact that at the point $m_1 = m_2 = 0$ the weights of the normalisable mode, and thus the mode itself, jump [see \eqref{eq:short15}]. 
We are going to investigate what happens exactly at this point in the next section.

\section{Partially massless gravity}\label{sec:6}

Above we calculated the LCFT new anomalies in GMG.
The evidence that there is an LCFT is based upon the analogous structures in TMG and NMG, where the LCFT conjecture is reasonably well-established for the rank 2 ``standard case''.
Here the LCFT has vanishing central charge and a logarithmic partner for the energy-momentum tensor.
However, we have seen above that GMG allows also for qualitatively different LCFTs, with non-vanishing central charges and no logarithmic partner for the energy-momentum tensor.
Thus, it is worthwhile to consider at least one example of such an LCFT in more detail and provide further evidence for the LCFT conjecture.

The example we consider in this section is called `partially massless gravity' (PMG).
PMG is NMG with the following tuning of parameters.
\eq{
2m^2\ell^2=\si
}{eq:PMG1}
This implies $\la=-1$ for the parameter in the NMG action \eqref{eq:NMG1}.
We consider exclusively the case $\si=-1$ here, and thus $m^2<0$.
We also set $\ell=1$ to reduce clutter.
PMG exhibits the following intriguing features, many of which were discovered already in \cite{Bergshoeff:2009aq} (see section 5.1 therein):
\begin{itemize}
 \item There is a unique AdS solution \eqref{eq:cg20} of PMG, which we take as the groundstate solution.
 \item The central charges of the dual CFT are given by 
\eq{
c_L=c_R=-\frac{3\ell}{G_N} 
}{eq:PMG2}
 \item The Breitenlohner--Freedman bound \cite{Breitenlohner:1982jf} is saturated.
\item The mass parameters $m_{1,2}$ in the linearised operators \eqref{eq:GMG3} vanish, so in Fig.~\ref{fig:2} PMG sits at the origin.
\item The two massive modes degenerate with each other at the critical point \eqref{eq:PMG1} and lead to a massive and a logarithmic massive mode.
 \item There is an enhancement of gauge symmetries at the linearised level that reduces the number of physical degrees of freedom to one.
\footnote{Another interesting jump in degrees of freedom occurs for the tuning $\si = 0$. Here the theory becomes conformal at the linearised level as spelled out in \cite{Deser:2009hb}.}
 \item The massive excitations become partially massless in the sense of Deser and Waldron \cite{Deser:1983mm,Deser:2001pe}. 
 \item The massive mode becomes pure gauge and its logarithmic partner constitutes a propagating bulk degree of freedom.
 \item The massive and logarithmic massive modes do not have a well-defined helicity.
 \item The Fefferman--Graham expansion of the massive mode becomes an ordinary power series in $e^\rho$ (possibly with subleading logarithms), rather than a fractional power series thereof.
 \item The Fefferman--Graham expansion of the logarithmic massive mode becomes an ordinary power series in $e^\rho$, with some leading logarithms.
 \item The normalisable and non-normalisable modes mix. The SL$(2,\mathbb{R})$ primaries have half integer weights, so both their differences and sums are integers. 
\end{itemize}
The last three properties follow directly from the analyses in \cite{Li:2008dq,Grumiller:2009mw}.
The LCFT interpretation is perfectly consistent with all the properties above.
We thus conjecture that the dual CFT of PMG is an LCFT with negative central charges \eqref{eq:PMG2}, no degeneration of the energy-momentum tensor, and a Jordan cell of rank 2 built by the massive mode $\psi^m$ and its logarithmic partner $\psi^{\rm log}$.

\subsection{Evidence for LCFT behaviour in PMG}

We discuss now further evidence that supports this conjecture.
For some of the checks it is useful to have an explicit expression for the massive primary\footnote{Here ``primary'' refers to SL$(2,\mathbb{R})$, as we have not checked the action of the Virasoro generators $L_{-n},\bar L_{-n}$. It would be interesting to fill this gap. We expect that all statements in this paragraph remain true when ``primary'' refers to the full Virasoro algebra.
The word ``weight'' refers to the quantum numbers $h,\bar h$, which in the CFT interpretation are not necessarily the weights, but rather a combination of weights and levels. For primaries what we call ``weights'' are indeed the conformal weights.} with weights $(h, \bar h)=(3/2, -1/2)$:
\eq{
\psi_{\mu\nu}^m = \frac{e^{-i\,3u/2+i\,v/2}}{\cosh\rho}\,\left(\begin{array}{lll}
                                             \;\psi_{uu}=\sinh^2\!\rho\;  & \;\psi_{uv}=0\; & \;\psi_{u\rho}=i\,\tanh\rho \;\\
                                             & \;\psi_{vv}=0\; & \;\psi_{v\rho}=0 \;\\
                                             & & \;\psi_{\rho\rho}=-\frac{1}{\cosh^2\!\rho}\;
                                            \end{array}\right)_{\!\!\!\mu\nu}
}{eq:PMG4}
Note that the primary \eqref{eq:PMG4} can be written as
\eq{
\psi_{\mu\nu}^m = \nabla_\mu\nabla_\nu\zeta - g_{\mu\nu}\,\zeta\qquad \zeta:=-\frac{e^{-i\,3u/2+i\,v/2}\,\sinh^2\!\rho}{2\cosh\rho}
}{eq:PMG4.5}
This is a consequence of the fact that massive modes are pure gauge due to partial masslessness \cite{Bergshoeff:2009aq}.
It is straightforward to show that all descendants of \eqref{eq:PMG4.5} are also pure gauge.\footnote{Descendants can be written as $\Lie_\xi \psi_{\mu\nu}^m=\nabla_\mu\nabla_\nu\eta-g_{\mu\nu}\eta$, with $\eta=\xi^\al\nabla_\al\zeta$, where $\xi$ is a Killing vector of AdS and $\Lie_\xi$ is the Lie derivative along this vector.}
Descendants are created by acting on the primary \eqref{eq:PMG4} with the ladder operators $L_+$ and $\bar L_+$ of SL$(2,\mathbb{R})_L\times\,$SL$(2,\mathbb{R})_R$.
The SL$(2,\mathbb{R})_L$ generators read 
\eq{
L_0 = i\partial_u \qquad L_\pm = ie^{\mp iu}\,\Big(\frac{\cosh{2\rho}}{\sinh{2\rho}}\partial_u-\frac{1}{\sinh{2\rho}}\partial_v \pm \frac i2 \partial_\rho\Big)
}{eq:cg21}
with algebra
$\big[L_0,L_{\pm}\big]=\pm L_{\pm}, \big[L_-,L_+\big]=2L_0$.
The SL$(2,\mathbb{R})_R$ generators $\bar L_0$, $\bar L_+$, $\bar L_-$ satisfy the same algebra and are given by \eqref{eq:cg21} with $u\leftrightarrow v$ and $L\leftrightarrow \bar L$.
The massive primary with weights $(h,\bar h)=(-1/2, 3/2)$ is obtained from \eqref{eq:PMG4} by exchanging the $u$- and $v$-components.

\begin{figure}
\begin{center}
\setlength{\unitlength}{1mm}
\begin{picture}(100,100)(-10,-10)
\thicklines
%
\matrixput(5,55)(10,0){3}(0,10){3}{\circle{2}} 
\matrixput(5,5)(10,0){3}(0,10){5}{\circle{2}} 
\matrixput(35,55)(10,0){5}(0,10){3}{\circle*{2}} 
\matrixput(55,5)(10,0){3}(0,10){3}{\circle{2}} 
\matrixput(35,35)(10,0){2}(0,10){2}{\circle{2}} 
\matrixput(55,35)(10,0){3}(0,10){2}{\circle*{2}} 
\matrixput(35,5)(10,0){2}(0,10){3}{\circle{2}} 
\dottedline{2}(-5,50)(85,50) 
\dottedline{2}(-5,30)(85,30) 
\dottedline{2}(30,-5)(30,85) 
\dottedline{2}(50,85)(50,-5) 
\put(40,-10){\vector(0,1){100}} 
\put(-10,40){\vector(1,0){100}} 
\put(40,95){\makebox(0,0){$h$}}
\put(95,40){\makebox(0,0){$\bar h$}}
\thinlines
%
\matrixput(6,5)(10,0){2}(0,10){5}{\vector(1,0){8}}
\matrixput(36,35)(10,0){4}(0,10){2}{\vector(1,0){8}}
\matrixput(36,5)(10,0){1}(0,10){3}{\vector(1,0){8}}
\matrixput(56,5)(10,0){2}(0,10){3}{\vector(1,0){8}}
\matrixput(6,55)(10,0){7}(0,10){3}{\vector(1,0){8}}
%
\matrixput(44,55)(-10,0){1}(0,10){3}{\vector(-1,0){8}}
\matrixput(74,55)(-10,0){2}(0,10){3}{\vector(-1,0){8}}
\matrixput(24,55)(-10,0){2}(0,10){3}{\vector(-1,0){8}}
\matrixput(74,35)(-10,0){2}(0,10){2}{\vector(-1,0){8}}
\matrixput(44,35)(-10,0){4}(0,10){2}{\vector(-1,0){8}}
\matrixput(74,5)(-10,0){7}(0,10){3}{\vector(-1,0){8}}
%
\matrixput(5,6)(10,0){8}(0,10){2}{\vector(0,1){8}}
\matrixput(5,36)(10,0){5}(0,10){1}{\vector(0,1){8}}
\matrixput(5,56)(10,0){8}(0,10){2}{\vector(0,1){8}}
\matrixput(55,26)(10,0){3}(0,10){3}{\vector(0,1){8}}
\matrixput(35,46)(10,0){2}(0,10){1}{\vector(0,1){8}}
%
\matrixput(5,74)(10,0){3}(0,-10){7}{\vector(0,-1){8}}
\matrixput(35,74)(10,0){5}(0,-10){2}{\vector(0,-1){8}}
\matrixput(35,44)(10,0){5}(0,-10){1}{\vector(0,-1){8}}
\matrixput(35,24)(10,0){5}(0,-10){2}{\vector(0,-1){8}}
\matrixput(35,34)(10,0){2}(0,10){1}{\vector(0,-1){8}}
%
\put(35,55){\circle{3}}
\put(55,35){\circle{3}}
\put(55,55){\circle{3}}
\end{picture}
\caption{Display of the ladder operators $L_+$ (up), $L_-$ (down), $\bar L_+$ (right) and $\bar L_-$ (left). Notation: $\odot=$~primaries, $\bullet=$~descendants, $\circ=$~other, $\dots=$~semi-permeable barrier.}
\label{fig:alg}
\end{center}
\end{figure}
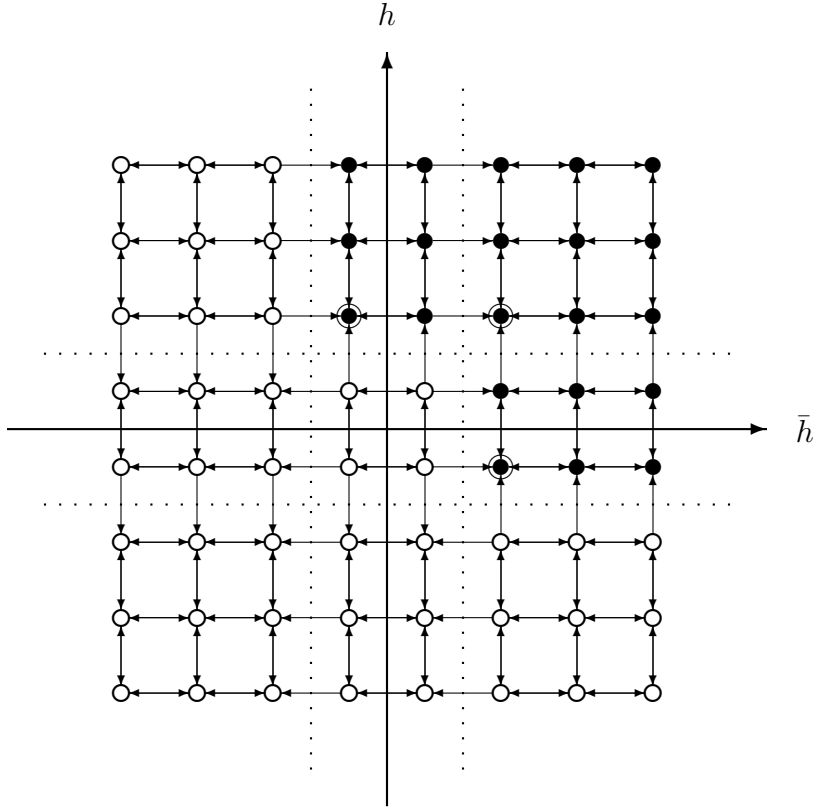

In Fig.~\ref{fig:alg} we depict all massive modes with half-integer weights and how they are connected by the SL$(2,\mathbb{R})$ ladder operators $L_\pm, \bar L_\pm$. 
In this picture filled circles refer to primaries and their descendants (a Verma module), while empty circles refer to all modes that are not descendants of a primary (the latter are usually non-normalisable modes).
Note that there are not only two primaries (and their complex conjugates) with the expected weights $(h, \bar h)=(3/2, -1/2)$, $(h, \bar h)=(1/2, - 3/2)$, but also an additional one with weights $(h, \bar h) = (3/2,3/2)$ that we denote by $\psi_{\mu\nu}^{0}$, where the superscript 0 refers to the vanishing angular momentum $h-\bar h=0$.
\eq{
\psi_{\mu\nu}^0 = \frac{e^{- i\,3/2\,(u+v)}}{\cosh^3\!\rho}\,\left(\begin{array}{lll}
                                             \;\psi_{uu}=1\;  & \;\psi_{uv}=\frac{3-\cosh{(2\rho)}}{2}\; & \;\psi_{u\rho}=- 3i\,\tanh\rho \;\\
                                             & \;\psi_{vv}=1\; & \;\psi_{v\rho}=- 3i\,\tanh\rho \;\\
                                             & & \;\psi_{\rho\rho}=\frac{4-2\cosh{(2\rho)}}{\cosh\!^2\rho}\;
                                            \end{array}\right)_{\!\!\!\mu\nu}
}{eq:newprimary}
Its existence is a special feature of PMG.
Interestingly, this additional primary arises as descendant of the other primaries. 
The new primary $\psi^0$ and its descendants fall off asymptotically like $e^{-\rho}$.
More specifically, we have for $h,\bar h\geq 3/2$:
\eq{
\psi_{uv}\sim e^{-\rho}\qquad \textrm{other\;components:\quad}\psi_{\mu\nu}\sim e^{-3\rho}
}{eq:added1} 
This is to be contrasted with the behaviour for $h\geq 3/2, \bar h\leq 1/2$:
\eq{
\psi_{uu}\sim e^\rho\qquad\psi_{uv}\sim e^{-\rho}\sim\psi_{u\rho}\qquad \textrm{other\;components:\quad}\psi_{\mu\nu}\sim e^{-3\rho}
}{eq:added2} 
We stress that for PMG there appears to be no separation into normalisable and non-normalisable modes: all modes grow asymptotically at most like $e^{\rho}$ and are thus compatible with the boundary conditions for fluctuations \eqref{eq:PMG8} below.
In the CFT interpretation\footnote{We thank Matthias Gaberdiel and Ivo Sachs for discussions on this interpretation.} the existence of the new primary $\psi^0$ means that the Verma module is reducible, i.e., there is a singular vector (or null vector) that generates a submodule of the whole Verma module. Such null vectors lead to differential equations that put restrictions on the higher correlators, so it is interesting that they occur automatically in PMG.

The logarithmic (quasi-)primary is obtained from the massive one \eqref{eq:PMG4} in complete analogy to \cite{Grumiller:2008qz}.
\eq{
\psi^{\rm log} = -\frac12\,\big(i(u+v)+\ln{\cosh^2\!{\rho}}\big)\,\psi^m
}{eq:PMG5}
Descendants are created by acting on the (quasi-)primary with the ladder operators $L_+$ and $\bar L_+$.
Acting with the operator that annihilates the massive mode on the logarithmic massive mode yields its logarithmic partner, the massive mode.
\eq{
\varepsilon_\mu{}^{\al\be}\nabla_\al \psi^{\rm log}_{\be\nu} = \psi^m_{\mu\nu}
}{eq:PMG6}
The algebraic discussion of the logarithmic modes parallels the one of the massive modes in the previous paragraph, except that ``primary'' should be replaced by ``quasi-primary'' when referring to the full Virasoro algebra, and the asymptotic behaviour changes logarithmically in $e^\rho$.
Acting on the pair $\psi^m(h=3/2,\bar h=-1/2)$ [see \eqref{eq:PMG4}] and $\psi^{\rm log}(h=3/2,\bar h=-1/2)$ [see \eqref{eq:PMG5}] with the Hamiltonian $H=L_0+\bar L_0$ and the angular momentum operator $J=L_0-\bar L_0$ yields 
\eq{ 
H \left(\begin{array}{c} \psi^{\rm log} \\ \psi^m
\end{array}\right) = \left(\begin{array}{c@{\quad}c}
1 & 1 \\
0 & 1
\end{array}\right) \left(\begin{array}{c} \psi^{\rm log} \\ \psi^m \end{array}\right)
}{eq:PGM10} 
and
\eq{ 
J \left(\begin{array}{c} \psi^{\rm log} \\ \psi^m
\end{array}\right) = \left(\begin{array}{c@{\quad}c}
2 & 0 \\
0 & 2
\end{array}\right) \left(\begin{array}{c} \psi^{\rm log} \\ \psi^m \end{array}\right)
}{eq:PGM11} 
The rank 2 Jordan cell in \eqref{eq:PGM10} 
provides another confirmation of the LCFT conjecture.

Further confirmation comes from the consideration of the asymptotic symmetry group, in analogy to \cite{Grumiller:2008es,Henneaux:2009pw}.
Let us consider asymptotic AdS
\eq{
\bar g_{\mu\nu}\extd x^\mu\extd x^\nu = \frac{\extd x^+\extd x^- + \extd y^2}{y^2}
}{eq:PMG7}
and allow for fluctuations $g=\bar g + h$ that preserve the following boundary conditions
\eq{
h_{\mu\nu}=\left(\begin{array}{lll}
       \;h_{++}={\cal O}\big(\frac{\ln y}{y}\big)\; & \;h_{+-}={\cal O}\big(1\big)\; & \;h_{+y}={\cal O}\big(\!\ln y\big)\; \\
       & \;h_{--}={\cal O}\big(\frac{\ln y}{y}\big)\; &  \;h_{-y}={\cal O}\big(\!\ln y\big)\; \\
       & & \;h_{yy} = {\cal O}\big(1\big)\;
      \end{array}\right)_{\!\!\!\mu\nu}
}{eq:PMG8}
These boundary conditions were discovered by Oliva, Tempo and Troncoso \cite{Oliva:2009ip}. They are weaker than Brown--Henneaux \cite{Brown:1986nw}, logarithmic \cite{Grumiller:2008es} or Giribet--Leston \cite{Giribet:2010ed} boundary conditions and 
differ also from Porfyriadis--Wilczek boundary conditions \cite{Porfyriadis:2010vg}.
In addition to the left- and right-moving boundary graviton excitations they allow for massive excitations like \eqref{eq:PMG4} as well as for logarithmic excitations \eqref{eq:PMG5}.
Vector fields $\xi$ that preserve the boundary conditions \eqref{eq:PMG8} when acting on the full metric $g=\bar g + h$ with the Lie derivative are given by
\begin{subequations}
 \label{eq:PMG9}
\begin{align}
 \xi^+ &= \eps^+(x^+) -  \frac{y^2}{2}\,\partial_-^2\eps^- + {\cal O}(y^{3}\ln y) \\
 \xi^- &= \eps^-(x^-) -  \frac{y^2}{2}\,\partial_+^2\eps^+ + {\cal O}(y^{3}\ln y) \\
 \xi^y &= \frac y2\,\big(\partial_+\eps^+ + \partial_-\eps^-\big) + {\cal O}(y^3)
\end{align}
\end{subequations}
The asymptotic symmetry group is then generated by the two functions $\eps^\pm(x^\pm)$, and thus naturally leads to two copies of the Witt algebra, which get centrally extended to two copies of the Virasoro algebra \cite{Oliva:2009ip} in full analogy to the seminal Brown--Henneaux example \cite{Brown:1986nw}.
Thus, PMG with the boundary conditions \eqref{eq:PMG8} is compatible with the conjecture that a dual CFT exists allowing for massive as well as logarithmic excitations.

In conclusion, we have provided the conjecture --- and evidence in its favour --- that the dual CFT of PMG is a specific LCFT with properties listed at the start of this section.
If the conjecture is true then PMG is a first example of a gravity dual for an LCFT where both central charges are negative and the energy-momentum tensor does not acquire a logarithmic partner.
It could be rewarding to perform further checks on the validity of our conjecture, like the calculation of correlators or partition functions, along the lines of Section \ref{sec:2}, or the calculation of the holographically renormalised Brown--York stress tensor, using the results of \cite{Hohm:2010jc}.

\subsection{New anomaly in PMG}

There is a subtlety in determining the new anomaly $b$ for PMG, which we
anticipated already after deriving the result \eqref{eq:mass6} for the
line $m_1 = m_2$. Note that $b$ changes sign abruptly as we pass the
point $m_1 = m_2 = 0$:
\eq{
b_{\rm PMG} = \mp \frac{3\ell \si}{G_N} 
}{eq:bPMG}
To interpret this, note that the weights of the normalisable modes
depend discontinuously on the point in the $(m_1, m_2)$--plane.
E.g., for the regions with the same sign of $m_i$ we have the weights
\begin{subequations}
\label{eq:regions}
\begin{align}
& (h, \bar{h}) =( \frac{3 + m_{1,2}\ell}{2},\frac{-1 + m_{1,2}\ell}{2}) \qquad m_i > 0\\
& (h, \bar{h}) =( \frac{-1 - m_{1,2}\ell}{2},\frac{3 - m_{1,2}\ell}{2}) \qquad m_i < 0
\end{align}
\end{subequations}
for the primaries of the normalisable modes. The modes correspond to
CFT operators having these same conformal weights.
It is clear that the result of a limiting construction as in Section
\ref{sec:4} will give different results depending on the region
from which the point $m_1 = m_2 = 0$ is approached: the
logarithmic operator will have different conformal weights!

Since the fall-off behaviour of all four modes in \eqref{eq:regions}
coincides at the PMG point, there is no {\it a priori} way to divide
them into (non-)normalisable modes. The situation is similar to the
case of a scalar field close to the Breitenlohner--Freedman bound:
there are two normalisable modes, and which one to consider as source
or operator is a matter of choice, see for instance the discussion in \cite{Witten:1998qj,Freedman:1998tz,Klebanov:1999tb,Witten:2001ua,Gubser:2002zh,Gubser:2002vv,Hartman:2006dy}.
Since the weights differ by an integer we have a situation resembling the case of resonant scalars, see \cite{Banados:2006de} and references therein.
In the present case this choice affects the sign
of the new anomaly \eqref{eq:bPMG}. The upper (lower) sign corresponds to the upper (lower) equation in \eqref{eq:regions}.

\section{Summary, generalisations and outlook} \label{sec:7}

We reviewed the status of the LCFT conjecture in 3-dimensional massive gravity theories in Sections \ref{sec:2} and \ref{sec:3}.
We then presented a short-cut to calculate new anomalies in gravity duals for logarithmic conformal field theories (LCFTs) in Section \ref{sec:4}. 
Exploiting this short-cut we derived in Section \ref{sec:5} results for new anomalies in generalised massive gravity (GMG) for the rank 2 standard case, \eqref{eq:news2}, the rank 3 standard case, \eqref{eq:three8}, and the rank 2 exotic case, \eqref{eq:mass6}.
``Standard'' refers to LCFTs with vanishing central charge where the energy-momentum tensor acquires a logarithmic partner.
``Exotic'' refers to LCFTs with non-vanishing central charge where massive modes degenerate with each other.
We discussed in some detail an exotic example of partially massless gravity (PMG) and found several intriguing features in Section \ref{sec:6}.
This theory is likely to provide the first gravity dual to an LCFT with non-vanishing central charges.

\newcommand{\Ric}{\slashed{R}}

We consider now generalisations to higher-derivative gravity theories with holographic $c$-theorem \cite{Sinha:2010ai,Paulos:2010ke,Sinha:2010pm}.
These theories are defined by the following action.\footnote{%
It is straightforward to add the gravitational Chern--Simons term to the action \eqref{eq:conclusion1}. 
This leads to the same linearised equations of motion as in GMG \eqref{eq:GMG2}, instead of \eqref{eq:conclusion3} below.
Another extension of NMG, Born--Infeld massive gravity \cite{Gullu:2010pc}, also is consistent with a holographic $c$-theorem, but does not have an LCFT dual for any regular choice of parameters.}
\eq{
S = \frac{1}{\ka^2}\,\int\extd^3x\sqrt{-g}\,\big[\sigma R-2\La+\sum_{nmk}\,\la_{nmk} R^n R_{(2)}^m R_{(3)}^k\Big]
}{eq:conclusion1}
The scalars $R_{(2)}=\Ric_{\mu\nu}\Ric^{\mu\nu}$ and $R_{(3)}=\Ric_{\mu\nu}\Ric^\mu_\al\Ric^{\nu\al}$ are quadratic and cubic curvature invariants constructed from the tracefree Ricci-tensor $\Ric_{\mu\nu}=R_{\mu\nu}-\frac13\,Rg_{\mu\nu}$.
The coupling constants $\la_{nmk}$ are not arbitrary, but restricted by the required existence of a holographic $c$-theorem.
To quadratic order these coupling constants match precisely the ones of NMG and the action \eqref{eq:conclusion1} reduces to \eqref{eq:NMG1} with $\La=\la m^2$.
To cubic and quartic order the coupling constants were first determined by Sinha \cite{Sinha:2010ai}.
To quintic order the coupling constants were first determined by Paulos \cite{Paulos:2010ke}, who gave also a prescription how to obtain them for arbitrary orders.
Since the action is algebraic in the Ricci tensor, the linearised equations of motion are no more than fourth order. Around an AdS background they take the same form as for NMG \eqref{eq:NMG2}, \eqref{eq:NMG3}. Below we use the notation of \cite{Paulos:2010ke}; note that $\ell=L/\sqrt{f_\infty}$, $c$ is the central charge and $\gamma$ a free parameter that determines the mass scale in \eqref{eq:short20} as $m^2_2\ell^2=1-c/(2f_\infty\gamma)$.
\eq{
\big(\nabla^2+\frac{2f_\infty}{L^2}\big)\big(\nabla^2+\frac{2f_\infty}{L^2}+\frac{c}{2L^2\gamma}\big)\psi_{\mu\nu} = 0
}{eq:conclusion3}
Tuning the coupling constants $\la_{nmk}$ can lead to the same kind of degenerations that we discussed for NMG, including the partially massless limit that leads to PMG.
Thus, the results of the present paper can be applied to appropriately tuned versions of \eqref{eq:conclusion1}.
More concretely, applying the short-cut of Section \ref{sec:4} we predict for these theories in the standard scenario of an LCFT with vanishing central charge the following value for the new anomaly.\footnote{Miguel Paulos has conjectured independently the result $b\propto\gamma$ \cite{Paulos:personal}.}
\eq{
b = 8 f_\infty \gamma
}{eq:conclusion2}
It would be excellent to confirm our prediction \eqref{eq:conclusion2} by a direct calculation of 2-point correlators in higher-derivative gravity theories with holographic $c$-theorem.

The research presented in this paper can be extended in a number of ways.
In TMG it would be interesting to extend the discussion of partition functions beyond 1-loop, based upon the concepts of Witten \cite{Witten:2007kt}, Maloney and Witten \cite{Maloney:2007ud} and the results in \cite{Grumiller:2010xv}.
This could also shed further light on chiral gravity \cite{Li:2008dq} as a truncation of the dual LCFT \cite{Grumiller:2008qz,Maloney:2009ck}.
In NMG a cross-check not performed yet is the calculation of 3-point correlators, and it could be rewarding to fill this gap.
While the structure of GMG makes it plausible that the LCFT conjecture is also true for GMG, it is not verified to the same degree as in TMG or NMG.
Thus, it would be interesting to calculate 2-, 3-point correlators and 1-loop partition functions in GMG in order to support/falsify the LCFT conjecture.
Also the 1-point function, the holographically renormalised Brown--York stress tensor, should be checked for finiteness, tracelessness and conservation. The same remarks apply to the special case of PMG that we studied in some detail.
More specifically, it will be interesting to check if the Brown--York stress tensor is finite for modes with the boundary behaviour \eqref{eq:added2}.

We have checked that in GMG for any values of the coupling constants not all central charges and energies of massive excitations \eqref{eq:mass3} can be strictly positive, thus confirming the expectations from TMG and NMG.
However, for the LCFT tunings of the coupling constants discussed in this work it is possible that all these quantities are non-negative.
In that case it is of interest to calculate the sign of the energy of logarithmic excitations. 
Our expectation is that this sign is negative, by analogy to the TMG calculation \cite{Grumiller:2008qz}.

Partial masslessness arises not only in PMG, but also in GMG\footnote{We thank Branislav Cvetkovi{\'c} for collaboration on this topic, and for explaining \cite{Cvetkovic:2010} to us.} as soon as one of the mass parameters $m_1$ and $m_2$ vanishes (i.e., along the axes in Fig.~\ref{fig:2}). Except for certain tunings ($m_i\ell = \pm 1$ or $\mu\to\infty$ where PMG is recovered) there are no degenerating modes. Therefore, the CFT duals of generic partially massless theories are not logarithmic. It would be interesting to unravel the properties of these ``Deser-Waldron-CFTs'', see e.g.~\cite{Dolan:2001ih,Tekin:2003np}.
Another intriguing loose end is the rank 3 scenario in GMG. In that case all left-moving {\em and} logarithmic excitations become zero norm states, so it is conceivable that a truncation to something like ``chiral gravity'' is possible. It could be rewarding to study such a truncation and to check if it parallels the chiral gravity story in TMG, or whether there are essential differences to it. 
Moreover, it would be interesting to expand the present analysis to the supergravity extensions of the present models put forward in \cite{Andringa:2009yc, Bergshoeff:2010mf}.

Finally, it may well be that the (non-unitary!) massive gravity theories considered in this work are gravity duals to LCFTs only in a certain limit, comparable to the large $N$, large 't~Hooft coupling limit on the CFT side and the supergravity limit on the AdS side in the canonical example of AdS/CFT \cite{Maldacena:1997re}.
It remains a challenge to either show that this is not the case or to find an adequate embedding of these theories in a more fundamental one, such as string theory. 

\acknowledgments

We thank Mohsen Alishahiha, Steve Carlip, Geoffrey Comp\`ere, Branislav Cvetkovi{\'c}, Stanley Deser, Sabine Ertl, Matthias Gaberdiel, Gaston Giribet, Marc Henneaux, Olaf Hohm, Roman Jackiw, Alex Maloney, Miguel Paulos, Jorgen Rasmussen, Ivo Sachs, Kostas \mbox{Skenderis}, Wei Song, Andy Strominger, Marika Taylor, Ricardo Troncoso, Balt van Rees and Dima Vassilevich for discussions.

DG, NJ and TZ are supported by the START project Y435-N16 of the Austrian Science Foundation (FWF) and by the FWF project P21927-N16. 
NJ acknowledges financial support from the Erwin-Schr\"odinger Institute (ESI) during the workshop ``Gravity in three dimensions''.

\appendix

\section{Doubly logarithmic mode}\label{app:double}

\newcommand{\id}{\extd}
\newcommand{\replaced}{L_x}

The asymptotic expansion of the logarithmic modes for arbitrary weights $h,\bar h$ was derived for TMG by expanding the equations of motion for the massive mode to first order in a small parameter $\eps = (1-m_1)/2$ \cite{Grumiller:2009mw}. (We set $\ell=1$ in this appendix.)
In a similar fashion we derive the doubly logarithmic mode, but going up to second order in $\eps$.
\begin{align}
 \psi^{\text{log}^2}=\frac{1}{2}\frac{\id^2\psi^M}{\id\varepsilon^2}\Big|_{\varepsilon=0}
\label{eq:app0}
\end{align}
Here $\psi^M$ is a solution of
\begin{align} \label{eommassive}
 \psi_{\mu\nu}^M +\frac{1}{1-2\eps}\,\varepsilon_\mu^{\phantom{\mu}\alpha\beta}\nabla_\alpha\psi_{\beta\nu}^M=0 
\end{align}
All such solutions were given explicitly in terms of hypergeometric functions in \cite{Grumiller:2009mw}.
The prefactor one half in \eqref{eq:app0} comes from our choice of normalisation, $\mathcal{D}^L\psi^{\text{log}^2}=-2\psi^{\text{log}}$.
We obtain the following asymptotic result for large values of $x=\cosh{(2\rho)}$
\begin{subequations}
 \label{eq:log2asy}
\begin{align}
 \psi^{\text{log}^2}_{vv}=&e^{-i(hu+\bar{h}v)}\Big[\frac{h\bar{h}+x}{2}\replaced^2
	-(h+\bar{h}+h\bar{h})\replaced +1+h+\bar{h}+h\bar{h}\Big] +\mathcal{O}\big(\frac{\ln^2x}{x}\big) \\
 \psi^{\text{log}^2}_{uv}=&e^{-i(hu+\bar{h}v)}\Big[\frac{1-h^2}{2}\replaced^2 
	-(1-3h)(1+h)\replaced+(1-7h)(1+h) \Big]+\mathcal{O}\big(\frac{\ln^2x}{x}\big) \\
 \psi^{\text{log}^2}_{uu}=&e^{-i(hu+\bar{h}v)}\frac{h(1-h^2)}{\bar{h}}\Big[\frac{1}{2}(\ln\frac{x}{4}-i(u+v))^2-2(\psi(h-1)+\psi(-\bar{h})-\frac{3}{2}+2\gamma)\nonumber\\ &\cdot (\ln\frac{x}{4}-i(u+v))
	+7+4\gamma(-3+2\gamma)+2(-3+4\gamma)(\psi(h-1)+\psi(-\bar{h}))\nonumber\\
	&+2(\psi(h-1)+\psi(-\bar{h}))^2-2(\psi'(h-1)-\psi'(-\bar{h})) \Big]+\mathcal{O}\big(\frac{\ln^2x}{x}\big) \label{eq:relevant} \\
 \psi^{\text{log}^2}_{u\rho}=&\mathcal{O}\big(\frac{\ln^2x}{x}\big) \\
 \psi^{\text{log}^2}_{v\rho}=&-i\,e^{-i(hu+\bar{h}v)}\Big[h\replaced^2-2(1+h)\replaced+2(1+h) \Big] +\mathcal{O}\big(\frac{\ln^2x}{x}\big) \\
 \psi^{\text{log}^2}_{\rho\rho}=&e^{-i(hu+\bar{h}v)}\frac{4}{x}\Big[\frac{1-2h^2}{2}\replaced^2-2(1-3h)(1+h)\replaced+2(1-7h)(1+h) \Big] +\mathcal{O}\big(\frac{\ln^2x}{x^2}\big)
\end{align}
\end{subequations}
with $\replaced:=\ln x +i(u+v)$.
Here $\psi$ is the digamma function and $\psi'$ its derivative.
Note that for 2-point correlators particularly the last line of \eqref{eq:relevant} is relevant, as it will produce the $\ln^2$-terms in the correlator \eqref{eq:log2corr}. 
(The first two lines of that equation will contribute only to terms that can be eliminated by a redefinition of the operator $\Osq$.)
One can check explicitly that the mode \eqref{eq:log2asy} obeys asymptotically
\eq{
 \mathcal{D}^L\mathcal{D}^L\mathcal{D}^L\,\psi^{\text{log}^2}=\mathcal{D}^L\mathcal{D}^L(-2\,\psi^{\text{log}})=
	\mathcal{D}^L(4\,\psi^L)=0 
}{eq:app1}
with the asymptotic expansions and normalisations of $\psi^{\rm log}$ and $\psi^L$ as given in \cite{Grumiller:2009mw}.


\begin{thebibliography}{10}

\bibitem{Gurarie:1993xq}
V.~Gurarie, ``{Logarithmic operators in conformal field theory},'' {\em Nucl.
  Phys.} {\bf B410} (1993) 535--549,
\href{http://www.arXiv.org/abs/hep-th/9303160}{{\tt hep-th/9303160}}.

\bibitem{Flohr:2001zs}
M.~Flohr, ``{Bits and pieces in logarithmic conformal field theory},'' {\em
  Int. J. Mod. Phys.} {\bf A18} (2003) 4497--4592,
\href{http://www.arXiv.org/abs/hep-th/0111228}{{\tt hep-th/0111228}}.

\bibitem{Gaberdiel:2001tr}
M.~R. Gaberdiel, ``{An algebraic approach to logarithmic conformal field
  theory},'' {\em Int. J. Mod. Phys.} {\bf A18} (2003) 4593--4638,
\href{http://www.arXiv.org/abs/hep-th/0111260}{{\tt hep-th/0111260}}.

\bibitem{Binder:1986zz}
K.~Binder and A.~P. Young, ``{Spin glasses: Experimental facts, theoretical
  concepts, and open questions},'' {\em Rev. Mod. Phys.} {\bf 58} (1986)
801--976.

\bibitem{Cardy:1999zp}
J.~L. Cardy, ``{Logarithmic Correlations in Quenched Random Magnets and
  Polymers},''
\href{http://www.arXiv.org/abs/cond-mat/9911024}{{\tt cond-mat/9911024}}.

\bibitem{RezaRahimiTabar:2000qr}
M.~Reza Rahimi~Tabar, ``{Quenched Averaged Correlation Functions of the Random
  Magnets},'' {\em Nucl. Phys.} {\bf B588} (2000) 630--637,
\href{http://www.arXiv.org/abs/cond-mat/0002309}{{\tt cond-mat/0002309}}.

\bibitem{Gurarie:1999yx}
V.~Gurarie and A.~W.~W. Ludwig, ``{Conformal algebras of 2D disordered
  systems},'' {\em J. Phys.} {\bf A35} (2002) L377--L384,
\href{http://www.arXiv.org/abs/cond-mat/9911392}{{\tt cond-mat/9911392}}.

\bibitem{Maldacena:1997re}
J.~M. Maldacena, ``{The large N limit of superconformal field theories and
  supergravity},'' {\em Adv. Theor. Math. Phys.} {\bf 2} (1998) 231--252,
\href{http://www.arXiv.org/abs/hep-th/9711200}{{\tt hep-th/9711200}}.
\item[]
O.~Aharony, S.~S. Gubser, J.~M. Maldacena, H.~Ooguri, and Y.~Oz, ``{Large N
  field theories, string theory and gravity},'' {\em Phys. Rept.} {\bf 323}
  (2000) 183--386,
\href{http://www.arXiv.org/abs/hep-th/9905111}{{\tt hep-th/9905111}}.

\bibitem{Grumiller:2008qz}
D.~Grumiller and N.~Johansson, ``{Instability in cosmological topologically
  massive gravity at the chiral point},'' {\em JHEP} {\bf 07} (2008) 134,
\href{http://www.arXiv.org/abs/0805.2610}{{\tt 0805.2610}}.

\bibitem{Deser:1982vy}
S.~Deser, R.~Jackiw, and S.~Templeton, ``Three-dimensional massive gauge
  theories,'' {\em Phys. Rev. Lett.} {\bf 48} (1982)
975--978.
%
``Topologically massive gauge
  theories,'' {\em Ann. Phys.} {\bf 140} (1982)
372--411.
%
{\em Erratum-ibid.} {\bf 185} (1988) 406.

\bibitem{Deser:1982sv}
S.~Deser, ``{Cosmological Topological Supergravity},'' in {\em Quantum Theory
  Of Gravity}, S.~M. Christensen, ed., pp.~374--381.
\newblock Adam Hilger, Bristol, 1984.
\newblock \href{http://www.arXiv.org/abs/Print-82-0692 (Brandeis)}{{\tt
  Print-82-0692 (Brandeis)}}.

\bibitem{Skenderis:2009nt}
K.~Skenderis, M.~Taylor, and B.~C. van Rees, ``{Topologically Massive Gravity
  and the AdS/CFT Correspondence},'' {\em JHEP} {\bf 09} (2009) 045,
\href{http://www.arXiv.org/abs/0906.4926}{{\tt 0906.4926}}.

\bibitem{Grumiller:2009mw}
D.~Grumiller and I.~Sachs, ``{AdS$_3$/LCFT$_2$ -- Correlators in Cosmological
  Topologically Massive Gravity},'' {\em JHEP} {\bf 03} (2010) 012,
\href{http://www.arXiv.org/abs/0910.5241}{{\tt 0910.5241}}.

\bibitem{Bergshoeff:2009hq}
E.~A. Bergshoeff, O.~Hohm, and P.~K. Townsend, ``{Massive Gravity in Three
  Dimensions},'' {\em Phys. Rev. Lett.} {\bf 102} (2009) 201301,
\href{http://www.arXiv.org/abs/0901.1766}{{\tt 0901.1766}}.

\bibitem{Bergshoeff:2009aq}
E.~A. Bergshoeff, O.~Hohm, and P.~K. Townsend, ``{More on Massive 3D
  Gravity},'' {\em Phys. Rev.} {\bf D79} (2009) 124042,
\href{http://www.arXiv.org/abs/0905.1259}{{\tt 0905.1259}}.

\bibitem{Grumiller:2009sn}
D.~Grumiller and O.~Hohm, ``{AdS$_3$/LCFT$_2$ - Correlators in New Massive
  Gravity},'' {\em Phys. Lett.} {\bf B686} (2010) 264--267,
\href{http://www.arXiv.org/abs/0911.4274}{{\tt 0911.4274}}.

\bibitem{Alishahiha:2010bw}
M.~Alishahiha and A.~Naseh, ``{Holographic Renormalization of New Massive
  Gravity},''
\href{http://www.arXiv.org/abs/1005.1544}{{\tt 1005.1544}}.

\bibitem{Deser:1983mm}
S.~Deser and R.~I. Nepomechie, ``Gauge invariance versus masslessness in de
  sitter space,'' {\em Ann. Phys.} {\bf 154} (1984)
396.

\bibitem{Deser:2001pe}
S.~Deser and A.~Waldron, ``{Gauge invariances and phases of massive higher
  spins in (A)dS},'' {\em Phys. Rev. Lett.} {\bf 87} (2001) 031601,
\href{http://www.arXiv.org/abs/hep-th/0102166}{{\tt hep-th/0102166}}.
%
``{Partial masslessness of higher spins in (A)dS},''
  {\em Nucl. Phys.} {\bf B607} (2001) 577--604,
\href{http://www.arXiv.org/abs/hep-th/0103198}{{\tt hep-th/0103198}}.

\bibitem{Sinha:2010ai}
A.~Sinha, ``{On the new massive gravity and AdS/CFT},'' {\em JHEP} {\bf 06}
  (2010) 061,
\href{http://www.arXiv.org/abs/1003.0683}{{\tt 1003.0683}}.

\bibitem{Paulos:2010ke}
M.~F. Paulos, ``{New massive gravity, extended},''
\href{http://www.arXiv.org/abs/1005.1646}{{\tt 1005.1646}}.

\bibitem{Sinha:2010pm}
A.~Sinha, ``{On higher derivative gravity, c-theorems and cosmology},''
\href{http://www.arXiv.org/abs/1008.4315}{{\tt 1008.4315}}.

\bibitem{Kraus:2005zm}
P.~Kraus and F.~Larsen, ``{Holographic gravitational anomalies},'' {\em JHEP}
  {\bf 01} (2006) 022,
\href{http://www.arXiv.org/abs/hep-th/0508218}{{\tt hep-th/0508218}}.

\bibitem{Li:2008dq}
W.~Li, W.~Song, and A.~Strominger, ``{Chiral Gravity in Three Dimensions},''
  {\em JHEP} {\bf 04} (2008) 082,
\href{http://www.arXiv.org/abs/0801.4566}{{\tt 0801.4566}}.

\bibitem{Grumiller:2008pr}
D.~Grumiller, R.~Jackiw, and N.~Johansson, ``{Canonical analysis of
  cosmological topologically massive gravity at the chiral point},'' in {\em
  {Fundamental Interactions - A Memorial Volume for Wolfgang Kummer}}.
\newblock World Scientific, 2009.
\newblock
\href{http://www.arXiv.org/abs/0806.4185}{{\tt 0806.4185}}.
\newblock

\bibitem{Carlip:2008qh}
S.~Carlip, ``{The Constraint Algebra of Topologically Massive AdS Gravity},''
  {\em JHEP} {\bf 10} (2008) 078,
\href{http://www.arXiv.org/abs/0807.4152}{{\tt 0807.4152}}.

\bibitem{Henneaux:2009pw}
M.~Henneaux, C.~Martinez, and R.~Troncoso, ``{Asymptotically anti-de Sitter
  spacetimes in topologically massive gravity},'' {\em Phys. Rev.} {\bf D79}
  (2009) 081502R,
\href{http://www.arXiv.org/abs/0901.2874}{{\tt 0901.2874}}.

\bibitem{Maloney:2009ck}
A.~Maloney, W.~Song, and A.~Strominger, ``{Chiral Gravity, Log Gravity and
  Extremal CFT},'' {\em Phys. Rev.} {\bf D81} (2010) 064007,
\href{http://www.arXiv.org/abs/0903.4573}{{\tt 0903.4573}}.

\bibitem{Ertl:2009ch}
S.~Ertl, D.~Grumiller, and N.~Johansson, ``{Erratum to `Instability in
  cosmological topologically massive gravity at the chiral point',
  arXiv:0805.2610},''
\href{http://www.arXiv.org/abs/0910.1706}{{\tt 0910.1706}}.

\bibitem{Carlip:2008jk}
S.~Carlip, S.~Deser, A.~Waldron, and D.~K. Wise, ``{Cosmological Topologically
  Massive Gravitons and Photons},'' {\em Class. Quant. Grav.} {\bf 26} (2009)
  075008,
\href{http://www.arXiv.org/abs/0803.3998}{{\tt 0803.3998}}.
``{Topologically Massive AdS
  Gravity},'' {\em Phys. Lett.} {\bf B666} (2008) 272--276,
\href{http://www.arXiv.org/abs/0807.0486}{{\tt 0807.0486}}.

\bibitem{Grumiller:2008es}
D.~Grumiller and N.~Johansson, ``{Consistent boundary conditions for
  cosmological topologically massive gravity at the chiral point},'' {\em Int.
  J. Mod. Phys.} {\bf D17} (2009) 2367--2372,
\href{http://www.arXiv.org/abs/0808.2575}{{\tt 0808.2575}}.

\bibitem{Li:2008yz}
W.~Li, W.~Song, and A.~Strominger, ``{Comment on 'Cosmological Topological
  Massive Gravitons and Photons'},''
\href{http://www.arXiv.org/abs/0805.3101}{{\tt 0805.3101}}.

\bibitem{Strominger:2008dp}
A.~Strominger, ``{A Simple Proof of the Chiral Gravity Conjecture},''
\href{http://www.arXiv.org/abs/0808.0506}{{\tt 0808.0506}}.

\bibitem{Brown:1986nw}
J.~D. Brown and M.~Henneaux, ``{Central Charges in the Canonical Realization of
  Asymptotic Symmetries: An Example from Three-Dimensional Gravity},'' {\em
  Commun. Math. Phys.} {\bf 104} (1986)
207--226.

\bibitem{Giribet:2008bw}
G.~Giribet, M.~Kleban, and M.~Porrati, ``{Topologically Massive Gravity at the
  Chiral Point is Not Unitary},'' {\em JHEP} {\bf 10} (2008) 045,
\href{http://www.arXiv.org/abs/0807.4703}{{\tt 0807.4703}}.

\bibitem{Compere:2010xu}
G.~Compere, S.~de~Buyl, and S.~Detournay, ``{Non-Einstein geometries in Chiral
  Gravity},''
\href{http://www.arXiv.org/abs/1006.3099}{{\tt 1006.3099}}.

\bibitem{Andrade:2009ae}
T.~Andrade and D.~Marolf, ``{No chiral truncation of quantum log gravity?},''
  {\em JHEP} {\bf 03} (2010) 029,
\href{http://www.arXiv.org/abs/0909.0727}{{\tt 0909.0727}}.

\bibitem{Gaberdiel:2007ve}
M.~R. Gaberdiel, ``{Constraints on extremal self-dual CFTs},'' {\em JHEP} {\bf
  11} (2007) 087,
\href{http://www.arXiv.org/abs/0707.4073}{{\tt 0707.4073}}.

\bibitem{Gaberdiel:2008pr}
M.~R. Gaberdiel and C.~A. Keller, ``{Modular differential equations and null
  vectors},'' {\em JHEP} {\bf 09} (2008) 079,
\href{http://www.arXiv.org/abs/0804.0489}{{\tt 0804.0489}}.

\bibitem{Grumiller:2010xv}
M.~R. Gaberdiel, D.~Grumiller, and D.~Vassilevich, ``{Graviton 1-loop partition
  function for 3-dimensional massive gravity},''
\href{http://www.arXiv.org/abs/1007.5189}{{\tt 1007.5189}}.

\bibitem{Liu:2009bk}
Y.~Liu and Y.-W. Sun, ``{Note on New Massive Gravity in $AdS_3$},'' {\em JHEP}
  {\bf 04} (2009) 106,
\href{http://www.arXiv.org/abs/0903.0536}{{\tt 0903.0536}}.

\bibitem{Liu:2009kc}
Y.~Liu and Y.-W. Sun, ``{Consistent Boundary Conditions for New Massive Gravity
  in $AdS_3$},'' {\em JHEP} {\bf 05} (2009) 039,
\href{http://www.arXiv.org/abs/0903.2933}{{\tt 0903.2933}}.

\bibitem{Liu:2009pha}
Y.~Liu and Y.-W. Sun, ``{On the Generalized Massive Gravity in $AdS_3$},'' {\em
  Phys. Rev.} {\bf D79} (2009) 126001,
\href{http://www.arXiv.org/abs/0904.0403}{{\tt 0904.0403}}.

\bibitem{RahimiTabar:1996ub}
M.~R. Rahimi~Tabar, A.~Aghamohammadi, and M.~Khorrami, ``{The logarithmic
  conformal field theories},'' {\em Nucl. Phys.} {\bf B497} (1997) 555--566,
\href{http://www.arXiv.org/abs/hep-th/9610168}{{\tt hep-th/9610168}}.

\bibitem{Khorrami:1997ci}
M.~Khorrami, A.~Aghamohammadi, and M.~R. Rahimi~Tabar, ``{Logarithmic conformal
  field theories with continuous weights},'' {\em Phys. Lett.} {\bf B419}
  (1998) 179--185,
\href{http://www.arXiv.org/abs/hep-th/9711155}{{\tt hep-th/9711155}}.

\bibitem{Rasmussen:2004gx}
J.~Rasmussen, ``{Logarithmic limits of minimal models},'' {\em Nucl. Phys.}
  {\bf B701} (2004) 516--528,
\href{http://www.arXiv.org/abs/hep-th/0405257}{{\tt hep-th/0405257}}.

\bibitem{Rasmussen:2004na}
J.~Rasmussen, ``{Jordan cells in logarithmic limits of conformal field
  theory},'' {\em Int. J. Mod. Phys.} {\bf A22} (2007) 67--82,
\href{http://www.arXiv.org/abs/hep-th/0406110}{{\tt hep-th/0406110}}.

\bibitem{Grumiller:2010rm}
D.~Grumiller and N.~Johansson, ``{Gravity duals for logarithmic conformal field
  theories},'' {\em J. Phys. Conf. Ser.} {\bf 222} (2010) 012047,
\href{http://www.arXiv.org/abs/1001.0002}{{\tt 1001.0002}}.

\bibitem{Zamolodchikov:1986gt}
A.~B. Zamolodchikov, ``{Irreversibility of the Flux of the Renormalization
  Group in a 2D Field Theory},'' {\em JETP Lett.} {\bf 43} (1986)
730--732.

\bibitem{Zojer:thesis}
T.~Zojer, ``{New anomalies of Generalized Massive Gravity},'' Master's thesis,
  {Vienna University of Technology}, 2010.

\bibitem{Breitenlohner:1982jf}
P.~Breitenlohner and D.~Z. Freedman, ``{Stability in Gauged Extended
  Supergravity},'' {\em Ann. Phys.} {\bf 144} (1982)
249.

\bibitem{Cvetkovic:2010}
M.~Blagojevic and B.~Cvetkovic, ``{Hamiltonian analysis of BHT massive
  gravity},''
\href{http://www.arXiv.org/abs/1010.2596}{{\tt 1010.2596}}.

\bibitem{Deser:2009hb}
S.~Deser, ``{Ghost-free, finite, fourth order D=3 (alas) gravity},'' {\em Phys.
  Rev. Lett.} {\bf 103} (2009) 101302,
\href{http://www.arXiv.org/abs/0904.4473}{{\tt 0904.4473}}.

\bibitem{Oliva:2009ip}
J.~Oliva, D.~Tempo, and R.~Troncoso, ``{Three-dimensional black holes,
  gravitational solitons, kinks and wormholes for BHT masive gravity},'' {\em
  JHEP} {\bf 07} (2009) 011,
\href{http://www.arXiv.org/abs/0905.1545}{{\tt 0905.1545}}.

\bibitem{Giribet:2010ed}
G.~Giribet and M.~Leston, ``{Boundary stress tensor and counterterms for
  weakened AdS$_3$ asymptotic in New Massive Gravity},'' {\em JHEP} {\bf 09}
  (2010) 070,
\href{http://www.arXiv.org/abs/1006.3349}{{\tt 1006.3349}}.

\bibitem{Porfyriadis:2010vg}
A.~P. Porfyriadis and F.~Wilczek, ``{Effective Action, Boundary Conditions, and
  Virasoro Algebra for AdS$_3$},''
\href{http://www.arXiv.org/abs/1007.1031}{{\tt 1007.1031}}.

\bibitem{Hohm:2010jc}
O.~Hohm and E.~Tonni, ``{A boundary stress tensor for higher-derivative gravity
  in AdS and Lifshitz backgrounds},'' {\em JHEP} {\bf 04} (2010) 093,
\href{http://www.arXiv.org/abs/1001.3598}{{\tt 1001.3598}}.

\bibitem{Witten:1998qj}
E.~Witten, ``{Anti-de Sitter space and holography},'' {\em Adv. Theor. Math.
  Phys.} {\bf 2} (1998) 253--291,
\href{http://arXiv.org/abs/hep-th/9802150}{{\tt hep-th/9802150}}.

\bibitem{Freedman:1998tz}
D.~Z. Freedman, S.~D. Mathur, A.~Matusis, and L.~Rastelli, ``{Correlation
  functions in the CFT($d$)/AdS($d+1$) correspondence},'' {\em Nucl. Phys.}
  {\bf B546} (1999) 96--118,
\href{http://www.arXiv.org/abs/hep-th/9804058}{{\tt hep-th/9804058}}.

\bibitem{Klebanov:1999tb}
I.~R. Klebanov and E.~Witten, ``{AdS/CFT correspondence and symmetry
  breaking},'' {\em Nucl. Phys.} {\bf B556} (1999) 89--114,
\href{http://www.arXiv.org/abs/hep-th/9905104}{{\tt hep-th/9905104}}.

\bibitem{Witten:2001ua}
E.~Witten, ``{Multi-trace operators, boundary conditions, and AdS/CFT
  correspondence},''
\href{http://www.arXiv.org/abs/hep-th/0112258}{{\tt hep-th/0112258}}.

\bibitem{Gubser:2002zh}
S.~S. Gubser and I.~Mitra, ``{Double-trace operators and one-loop vacuum energy
  in AdS/CFT},'' {\em Phys. Rev.} {\bf D67} (2003) 064018,
\href{http://www.arXiv.org/abs/hep-th/0210093}{{\tt hep-th/0210093}}.

\bibitem{Gubser:2002vv}
S.~S. Gubser and I.~R. Klebanov, ``{A universal result on central charges in
  the presence of double-trace deformations},'' {\em Nucl. Phys.} {\bf B656}
  (2003) 23--36,
\href{http://www.arXiv.org/abs/hep-th/0212138}{{\tt hep-th/0212138}}.

\bibitem{Hartman:2006dy}
T.~Hartman and L.~Rastelli, ``{Double-trace deformations, mixed boundary
  conditions and functional determinants in AdS/CFT},'' {\em JHEP} {\bf 01}
  (2008) 019,
\href{http://www.arXiv.org/abs/hep-th/0602106}{{\tt hep-th/0602106}}.

\bibitem{Banados:2006de}
M.~Banados, A.~Schwimmer, and S.~Theisen, ``{Remarks on resonant scalars in the
  AdS/CFT correspondence},'' {\em JHEP} {\bf 09} (2006) 058,
\href{http://www.arXiv.org/abs/hep-th/0604165}{{\tt hep-th/0604165}}.

\bibitem{Gullu:2010pc}
I.~Gullu, T.~C. Sisman, and B.~Tekin, ``{Born-Infeld extension of new massive
  gravity},''
\href{http://www.arXiv.org/abs/1003.3935}{{\tt 1003.3935}}.

\bibitem{Paulos:personal}
M.~Paulos.
\newblock personal communication.

\bibitem{Witten:2007kt}
E.~Witten, ``{Three-Dimensional Gravity Revisited},''
\href{http://www.arXiv.org/abs/0706.3359}{{\tt 0706.3359}}.

\bibitem{Maloney:2007ud}
A.~Maloney and E.~Witten, ``{Quantum Gravity Partition Functions in Three
  Dimensions},''
\href{http://www.arXiv.org/abs/0712.0155}{{\tt 0712.0155}}.

\bibitem{Dolan:2001ih}
L.~Dolan, C.~R. Nappi, and E.~Witten, ``{Conformal operators for partially
  massless states},'' {\em JHEP} {\bf 10} (2001) 016,
\href{http://www.arXiv.org/abs/hep-th/0109096}{{\tt hep-th/0109096}}.

\bibitem{Tekin:2003np}
B.~Tekin, ``{Partially massless spin-2 fields in string generated models},''
\href{http://www.arXiv.org/abs/hep-th/0306178}{{\tt hep-th/0306178}}.

\bibitem{Andringa:2009yc}
R.~Andringa {\em et al.}, ``{Massive 3D Supergravity},'' {\em Class. Quant.
  Grav.} {\bf 27} (2010) 025010,
\href{http://www.arXiv.org/abs/0907.4658}{{\tt 0907.4658}}.

\bibitem{Bergshoeff:2010mf}
E.~A. Bergshoeff, O.~Hohm, J.~Rosseel, E.~Sezgin, and P.~K. Townsend, ``{More
  on Massive 3D Supergravity},''
\href{http://www.arXiv.org/abs/1005.3952}{{\tt 1005.3952}}.

\end{thebibliography}

\providecommand{\href}[2]{#2}\begingroup\raggedright\endgroup

\end{document}